\documentclass[pre,twocolumn,showpacs]{revtex4-1}

\usepackage{graphicx}
\usepackage{amsmath,amssymb}
\usepackage{bm}
\usepackage{cancel}
\usepackage{color}
\usepackage{hyperref}

\newcommand{\calA}{\mathit{A}}
\newcommand{\calG}{\mathit{G}}
\newcommand{\calL}{\mathit{L}}
\newcommand{\calP}{\mathcal{P}}
\newcommand{\1}{\eta^{(1)}}
\newcommand{\2}{\eta^{(2)}}
\newcommand{\3}{\eta^{(3)}}
\newcommand{\ei}{\eta^{(i)}}

\newcommand{\eq}{\text{eq}}
\newcommand{\auno}{a^{(1)}}
\newcommand{\atres}{a^{(3)}}
\newcommand{\ai}{a^{(i)}}
\newcommand{\guno}{g^{(1)}}
\newcommand{\gtres}{g^{(3)}}
\newcommand{\gi}{g^{(i)}}
\newcommand{\sgn}{\text{sgn}}

\graphicspath{{./}{figuras_pdf/}{figures_published_version/}}

\begin{document}
\title{Theory of force-extension curve for modular proteins and DNA hairpins}
\author{L.L. Bonilla$^1$, A. Carpio$^2$, A. Prados$^3$}
\affiliation {$^1$G. Mill\'an Institute, Fluid Dynamics, Nanoscience and Industrial
Mathematics, Universidad Carlos III de Madrid, 28911 Legan\'es, Spain}
\affiliation{$^2$ Departamento de Matem\'atica Aplicada, Universidad Complutense de Madrid, 28040 Madrid, Spain}
\affiliation{$^3$ F\'{\i}sica Te\'{o}rica, Universidad de Sevilla,
Apartado de Correos 1065, E-41080, Sevilla, Spain}

\date{\today}
\begin{abstract}
  We study a model describing the force-extension curves of modular
  proteins, nucleic acids, and other biomolecules made out of several
  single units or modules. At a mesoscopic level of
  description, the configuration of the system is given by the
  elongations of each of the units.  The system free energy includes a
  double-well potential for each unit and an elastic nearest neighbor
  interaction between them. Minimizing the free energy yields the
  system equilibrium properties whereas its dynamics is given by
  (overdamped) Langevin equations for the elongations, in which
  friction and noise amplitude are related by the
  fluctuation-dissipation theorem. Our results, both for the
  equilibrium and the dynamical situations, include analytical and
  numerical descriptions of the system force-extension curves under
  force or length control, and agree very well with actual experiments
  in biomolecules. Our conclusions also apply to other physical
  systems comprising a number of metastable units, such as storage
  systems or semiconductor superlattices.
\end{abstract}
\maketitle 

\section{Introduction}
\label{sec:0}

Nowadays technological advances allow manipulation of single molecules
with sufficient precision to study many mechanical, kinetic and
thermodynamic properties thereof. Recent reviews of techniques used
and results obtained in single-molecule experiments (SMEs) can be
found in Refs.~\cite{rit06jpcm,KyL10,MyD12}. In these
  experiments, typically the force applied to pull the biomolecule {is}
  recorded as a function of its end-to-end distance, {thereby producing a}
force-extension curve (FEC). This FEC
  characterizes the molecule elasticity and provides
information about its processes of folding and unfolding
\cite{smi96,car99,lu99,fis00,lip01,bus03,man05,cao08}. In the
  following, the end-to-end distance of the biomolecule is referred to
  as the total length \cite{length}. The force--extension curves are
different depending on whether the total length or the force are
controlled. When the total length of the protein is used as a control
parameter (length-control), the {unfolding} transition is accompanied
by a drop in the measured force and a sawtooth pattern is the typical
force--extension curve \cite{fis00,car99,lip01,bus03,lip02}. When the
force is the control parameter (force-control), unfolding of several
or all single protein domains may occur at a constant value of the
force \cite{hug10}. Other questions are related to the rate at which
the control parameter (length or force) sweeps the force--extension
curve: depending on the loading rate, stochastic jumps between folded
and unfolded protein states may be observed
\cite{rit06jpcm,lip01,man05,lip02,hug10}.

The analysis of the force vs. extension curves provides valuable
information about the polyprotein, the DNA or the RNA hairpin. Let us
consider atomic force microscope (AFM) experiments in which a
modular protein comprising a number of identical folds (modules or
units) is pulled at a certain rate (length-control) 
\cite{car99,fis00}. The typical value of the force $F_{c}$ at which
the unfolding takes place is related to the mechanical stability of
the units: a larger value of the force is the signature of higher
stability. Nevertheless, it should be stressed that the unraveling of
a domain is a stochastic event and occurs for forces within a certain
range.  A second feature of the sawtooth FEC is the spacing between
consecutive force peaks. This spacing is directly related to the
difference of length between the folded and unfolded configurations of
one unit.  This is the reason that the peaks of the FEC of
artificially engineered modular proteins are regularly spaced. A
typical example is I27$_8$, composed of eight copies of immunoglubulin
domain 27 from human cardiac titin.  The spacing between peaks for
this protein is $28.4\pm 0.3$nm at an unfolding force of $204\pm 26$pN
\cite{car99,fis00}. This length increment is found by fitting several
peaks of the FEC with the worm-like chain (WLC) model of polymer
elasticity \cite{bus94,mar95}.  More recently, force-controlled AFM
experiments with a I27 single-module protein have been reported
\cite{ber10,ber12}. These experiments provide data free from the
module to module variations that even an artificially engineered
polyprotein has. Berkovich et al.\/ have interpreted their results
using a simple Langevin equation model that includes an effective
potential with two minima for a range of the applied force
\cite{ber10}.

The thermodynamics of pulling experiments is well established under
both force and length control. For controlled force, the relevant
thermodynamic potential is a Gibbs-like free energy, whereas for
controlled length it is a Helmholtz-like free energy
\cite{KSyB03,man05,RByK06}. Interestingly, the sawtooth structure of
the FEC of biomolecules is already present at equilibrium, as shown
very recently in a simple model with a Landau-like free energy
\cite{pra13}.  However, the control parameter in real experiments with
biomolecules (force or length) changes usually with time at a finite
rate \cite{rit06jpcm,MyD12,fis00,lip01,lip02,man05,hug10,HBFSByR10}.
Knowledge of these dynamical situations is
not as complete as in the equilibrium case. Under force control, we
can write a Langevin equation (or the associated Fokker-Planck
equation) in which noise amplitude and effective friction are linked
by a fluctuation-dissipation relation, as done in
Refs.~\cite{ber10,RRyV05}. On the other hand, under length control,
the situation is more complex: the force is no longer a given function
of time but an unknown that must be calculated by imposing the length
constraint. This has lead to the proposal of simple \textit{dynamical
  algorithms} such as the quasi-equilibrium algorithm of
Ref.~\cite{man05}. While being successful in reproducing
experimentally observed behavior, these algorithms do not correspond
to the integration of well-defined evolution equations.

In some cycling experiments, the biomolecule is switched
  between the folded and unfolded configurations, at a certain
  switching rate \cite{rit06jpcm,lip01,lip02,man05,hug10,HBFSByR10}.
After Liphardt et al., we call such a process a
  stretching/relaxing or an unfolding/refolding cycle
  \cite{lip01,lip02}. The unfolding typically occurs at a force
$F_{u}$ that is larger than the refolding force $F_{r}$. Therefore,
some hysteresis is present and, moreover, the unfolding (refolding)
force typically increase (decrease) with the pulling rate. A
reversible curve in which $F_{u}=F_{r}=F_{c}$ is only observed for a
small enough rate. Some authors have claimed that this is a signature
of irreversible non-equilibrium behavior and thus used these
experiments to test non-equilibrium fluctuation theorems
\cite{lip02,PKTyS03,CRJSTyB05,HyS10}. On the other hand, for a simple
model for which only the force-controlled situation could be analyzed
\cite{pra12}, it has been found that the observed behavior in
biomolecules can be understood as the system sweeping a certain part
of the metastable equilibrium region of the FEC that surrounds
$F_{c}$. In this way, the system is exploring metastable minima of the
system free energy landscape.  One of the main goals of this work is
to determine if this physical picture also holds for length-controlled
experiments.

In this paper, we add two important ingredients of real biomolecule
pulling experiments to a simple model with independent domains and
Landau-like free energy whose equilibrium analysis is given in
Ref. \cite{pra13}. We add: (i) dynamical effects and
(ii) interacting units. Dynamical effects are introduced by means of
Langevin or Fokker-Planck equations, both under force and, most
interestingly, length control. Therefrom, we can carry out a
systematic investigation of the dynamical FEC, when the
control parameter (force or length) is varied at a finite rate. The
simplest way to introduce interaction between modules is via a
harmonic potential trying to drive them to global equilibrium. In
this way, the creation of \textit{bubbles}, that is, regions of
unfolded modules inside regions of folded ones, has a free energy
cost. This is expected to be most relevant for systems in which the
unfolding/refolding of units is mainly sequential, as in the unzipping
of DNA hairpins \cite{rit06jpcm}. Interestingly, the complex and
force-sensitive behavior of polyproteins observed in force-clamp
experiments has been recently explained by sequential
unfolding \cite{BCyP14}.

The main ingredients of our model are bistability of protein modules
and, in the length-controlled case, a global constraint that
introduces a long-range interaction among modules. These features are
quite general in physics, as they appear in many different fields. For
instance, many particle storage systems such as the storage of lithium
in multi-particle electrodes of rechargeable lithium-ion batteries
\cite{dre10,dre11}, air storage in interconnected systems of rubber
balloons \cite{dre11cmt}, or voltage biased weakly coupled
semiconductor superlattices
\cite{GHMyP91,RTGyP02,BGr05,BT10,BHGyT12}. Throughout the paper, the
analogies and differences that arise in these different physical
situations will be discussed.

The rest of the paper is as follows. The model we use is described in
Section \ref{model}, in which we write down both the Langevin and the
Fokker-Planck equations in Secs.~\ref{dynamics} and \ref{F-P},
respectively.  In Section \ref{ideal_chain}, we investigate an ideal
modular protein comprising many identical, non-interacting, units. In
Sec.~\ref{landau_free_energy}, we show that the equilibrium FEC
corresponding to our Landau-like double-well free energy has multiple
branches. Statistical mechanics considerations determine the stability
of the equilibrium branches for: (a) force-control in
Sec.~\ref{cont_force_eq}, and (b) length-control in
Sec.~\ref{cont_length_eq}.  We also consider dynamical situations when
the control parameter (either force or length) varies at a finite
rate. Section \ref{real_chain} deals with a real chain, in which the
nearest neighbor modules interact via an extra harmonic term. First,
we study the equilibrium situation in Sec.~\ref{eq_real}, in which we
show that the size of the branches is reduced, as compared to the
ideal case.  Sections \ref{dynamics_T=0}, \ref{quenched_disorder}, and
\ref{thermal_noise} analyze the changes that the dynamics brings to
the equilibrium picture by considering deterministic dynamics,
quenched disorder and finite temperature dynamics (thermal noise),
respectively. Final remarks are made in Section
\ref{conclusions}. Appendix A explains unfolding and refolding under
length control using a more realistic potential, whereas Appendices B
and C deal with some technical aspects not covered in the main text.

\setcounter{equation}{0}

\section{Model}
\label{model}

To be specific, let us consider AFM {unfolding} of modular proteins:
They are stretched between the tip of the microscope cantilever and a
flat, gold-covered substance (platform), whose position is externally
controlled. The forces acting on the molecule bend the cantilever
which, in turn, determines the applied force with pN precision. See
Fig.~1 of Refs.~\onlinecite{MyD12} or ~\onlinecite{fis00} for an idealized situation. In
force-controlled experiments with a single module protein, the free
energy of an extending protein comprises at least two distinct
components, an entropic term that accounts for chain elasticity and an
enthalpic component that includes the short-range interactions arising
between the neighboring amino acids as the protein contracts
\cite{ber10,ber12}. In a certain force range, these two components
cause the single-module free energy to have two minima,
corresponding to the folded and unfolded states of the
  domain \cite{ber10}.

Let us consider a system comprising $N$ modules. The $j$-th module extends from $x_j$ to $x_{j+1}$, so
that its extension is $\eta_j= x_{j+1}-x_j$, $j=1,\ldots,N$. The
configuration $\bm{\eta}=\{\eta_j\}$ defines the polyprotein state at
a mesoscopic level of description. When isolated, the free energy of
the $j$-th unit is $a(\eta_j;Y,\delta_j)$, a double-well potential
whose minima correspond to the folded and unfolded states discussed
above.  $Y$ is the set of relevant intensive parameters, like the
temperature $T$ and the pressure $p$ of the fluid (thermal bath)
surrounding our system. The parameter $\delta_j$ accounts for the
slight differences from unit to unit: $\delta_j=0$, $\forall j$, if
all units are identical and thus no \textit{quenched disorder} is
present in the system.

As part of the tertiary structure of the polyprotein, modules are
weakly interconnected by linkers in a structure-dependent way
\cite{HyD12}. It seems reasonable that this weak interaction acts on
the unfolding/refolding time scale and tries to bring the extensions
of the modules to a common value, corresponding to global mechanical
equilibrium. For the sake of simplicity, we model the linkers as
harmonic springs. Thus the system free energy $\calA$ for a given
configuration of module extensions $\bm{\eta}$ is
\begin{equation}\label{1.1}
  \calA(\bm{\eta};Y)=\sum_{j=1}^N a(\eta_j;Y,\delta_j)+\sum_{j=2}^{N} \frac{k_j(Y)}{2}  (\eta_j-\eta_{j-1})^2.
\end{equation}
If all the linkers are identical, $k_j=k$ for all $j=2,\ldots,N$, and
the elastic constants may depend on the intensive parameters.  The
length $\calL$ of a polyprotein in a configuration $\bm{\eta}$ is
\begin{equation}
\calL(\bm{\eta})=\sum_{j=1}^N \eta_j. \label{1.2b}
\end{equation}

The experiments are carried out at either force-controlled or
length-controlled conditions.  Firstly, we analyze case (i), in which
a certain external force $F=F(t)$ is applied to the ends of the
protein or DNA hairpin. For a detailed discussion of how this is
achieved in real experiments, see for example Refs.~\cite{hug10}
(ch.~6) and \cite{FLMHHyR11} for the optical tweezers case, and
\cite{OHCyF01,FyL04} for the AFM case.  In our simplified theoretical
approach, we only have to add a term
\begin{equation}\label{eq:deltaU_force}
\Delta U_{\text{fc}}(\bm{\eta};F)=-F L(\bm{\eta})=F x_1 - F x_{N+1}
\end{equation}
to the free energy $\calA(\bm{\eta})$. In this way, we obtain a Gibbs
free energy
$\calG(\bm{\eta};Y,F)=\calA(\bm{\eta},Y)+\Delta U_{\text{fc}}=\calA(\bm{\eta},Y)-F \calL(\bm{\eta})$,
\begin{subequations}\label{1.2}
\begin{equation}
 \calG(\bm{\eta};Y,F)= \sum_{j=1}^N  g(\eta_j;Y,F,\delta_j) +\sum_{j=2}^{N} \frac{k_j(Y)}{2}  
  (\eta_j-\eta_{j-1})^2, \label{1.2a}
\end{equation}
\begin{equation}
 g(\eta_j;Y,F,\delta_j)=a(\eta_j;Y,\delta_j)-F
  \eta_j.
\end{equation}
\end{subequations}
Note that we are not taking into account the limited bandwidth of the
feedback device that controls force in real experiments; we are
assuming that the desired force program $F(t)$ is perfectly
implemented. Secondly, we investigate the length-controlled situation,
case (ii). For a schematic representation of the experimental
  situation, see for instance Fig.~1 of
Ref.~\onlinecite{MyD12}: The length $L(t)$ between the base
of the cantilever and the platform is the externally
controlled {quantity}. {On the other hand,} the
cantilever tip deflects a certain distance $\Delta x$ from its base,
such that $\Delta x+\calL(\eta)=L(t)$. If the stiffness (spring
constant) of the cantilever is $\chi_{\text{lc}}$, we have an extra
harmonic term in the potential
$U_{\text{lc}}=\chi_{\text{lc}}(\Delta x)^{2}/2$, that is,
\begin{equation}\label{eq:length-control}
  \Delta U_{lc}(\bm{\eta};L)=\frac{\chi_{\text{lc}}}{2}\left[L(\bm{\eta})-L(t)\right]^{2},
\end{equation}
 Therefore, an extra force
$\Delta F_{lc}=-\chi_{\text{lc}}[L(\bm{\eta})-L(t)]$ acts over each
unit, trying to keep the polyprotein length equal to $L(t)$: The larger $ \chi_{\text{lc}}$, the better the length-control
is, as expected on intuitive grounds and explicitly shown in
\cite{PKTyS03}. In this paper, with the exception of Appendix A, we
assume perfect length control, that is, we consider the limit
$ \chi_{\text{lc}}\to\infty$ that implies $L(\bm{\eta})-L(t)\to 0$
over the time evolution of the units and a finite value of the
corresponding extra force $\Delta F_{lc}$.  In other words,
$\Delta F_{lc}$ tends to a limiting value $F$ that depends on the
prescribed length $L(t)$. This unknown value is the force required
to attain the total length $L(t)$, and {it} has to be calculated by
imposing the constraint $\sum_{i}x_{i}=L$. The effect of this limit on
the relevant thermodynamic potential for the length-controlled case
$A+\Delta U_{lc}$ shall be discussed in the section on Fokker-Planck
description of the dynamics.

Finally, we would like to stress that the present model has some
similarities with the more complicated one proposed by Hummer and
Szabo for the unfolding of polyproteins several years ago, see
Appendix C of Ref.~\cite{HyS03}. In addition to the module extensions
$\eta_j$, these authors consider the module centers of mass $r_j$ as
independent unknowns. These variables interact through a
quadratic potential that yields a linear restoring force whenever
$r_{j+1}-r_j$ departs from $\frac{\eta_{j+1}+\eta_j}{2}$. The site
potential for the module extensions is the sum of a WLC potential and
a harmonic potential \cite{HyS03}, instead of the double-well
potential we consider in the main text or the asymmetric potential we
consider in Appendix A. Moreover, Hummer and Szabo introduce
a WLC linker that connects the polyprotein with the length-controlling
device, which is absent in our model.

\subsection{Langevin dynamics}\label{dynamics}

The extensions $\eta_j$ obey coupled Langevin equations with the
appropriate thermodynamic potential. The friction coefficient and the amplitude of the white
noise are related by a fluctuation-dissipation theorem. The
\textit{source} for both the friction and the stochastic force is the
fluid the modules are immersed in, which is assumed to remain in
equilibrium at temperature $T$.  We assume that the modules' inertia
can be neglected and thus their evolution equations are overdamped, 
\begin{subequations}\label{langevin_eqs}
\begin{eqnarray}
 \gamma_j\dot{\eta}_j &=& F -\frac{\partial}{\partial \eta_j} \calA(\bm{\eta};Y) +\sqrt{2 T\gamma_j}\, \xi_j(t) ,
\label{1}
\end{eqnarray}
\begin{equation}
\langle\xi_j(t)\rangle\!=\!0, \, \langle\xi_j(t)\xi_l(t')\rangle\!=\!\delta_{jl}\,\delta(t-t'), \, j\!=\!1,\ldots,N.   \label{1.3}
\end{equation}
\end{subequations}
Here $\gamma_j$ is the friction coefficient for the $j$-th module, and
we measure the temperature in units of energy ($k_{B}=1$).  In
general, the friction coefficients $\gamma_{j}$ may depend on the
system configuration $\bm{\eta}$, if hydrodynamic interactions play a
significant role in the considered {unfolding} scenario. For the sake
of simplicity, we do not consider this possibility in the present
paper. In this respect, it is interesting to remark that, in
  the more complicated model of Ref.~\cite{HyS03}, module centers of mass and
  extensions satisfy Langevin equations with different
  extension-dependent diffusion coefficients.

Our presentation of the model above implies that the Langevin
  equations \eqref{langevin_eqs} are valid both in force-controlled
  and length-controlled experiments, but (i) in force-controlled
experiments, $F=F(t)$ is the known force program,
whereas (ii) in length-controlled ones we have
$\calL(\bm{\eta})=L(t)$. We differ the discussion on the
  experimental situation with an ``imperfect'' length control (because
  of the finite value of the stiffness $\chi_{\text{lc}}$ of the
  device controlling the length) to the next section on the equivalent
  Fokker-Planck description of the dynamics. For perfect length
  control, $F(t)$ is determined by imposing the constraint
$L(\bm{\eta})=L$, which yields
\begin{subequations}
\begin{eqnarray}
    F=\frac{\gamma}{N}\!\left( \frac{dL}{dt}+\sum_{j=1}^N \frac{1}{\gamma_j} \frac{\partial\calA(\bm{\eta};Y)}{\partial \eta_j} -\sum_{j=1}^N\sqrt{\frac{2T}{\gamma_j}}\xi_j\right)\!\!.
\label{1.3c}
\end{eqnarray}
\begin{equation}\label{1.3b}
  \gamma^{-1}=\frac{1}{N} \sum_{j=1}^N \gamma_j^{-1}.
\end{equation}
\end{subequations}
The parameter $\gamma$ is an average friction coefficient. In the case of identical units, $\gamma_j=\gamma$, $\forall j$. We split $F$ in two terms, a ``macroscopic term'' $F_{FP}$ and a ``fluctuating term''  $\Delta F$, as follows:
\begin{subequations}
\begin{equation}
    \label{eq:splitF}
F=F_{\text{FP}}+\Delta F,
    \end{equation}
\begin{eqnarray}
  \label{eq:Fexp}
F_{\text{FP}} & = & \frac{\gamma}{N}\left[ \frac{dL}{dt}+\sum_{j=1}^N
  \frac{1}{\gamma_j} \frac{\partial\calA(\bm{\eta};Y)}{\partial
    \eta_j} \right], \\
  \label{eq:DeltaF} \Delta F & = & -\frac{\gamma}{N} \sum_{j=1}^N\sqrt{\frac{2T}{\gamma_j}}\xi_j.
\end{eqnarray}
\end{subequations}
We prove in Sec.~\ref{F-P} that $F_{FP}$ is the force appearing in the
flux term of the Fokker-Plack equation.  Note that for any
$N$, $\langle \Delta F\rangle=0$ and then
$\langle F\rangle=\langle F_{\text{FP}}\rangle$. Furthermore,
  $\Delta F$ is a sum of Gaussian variables, and thus its statistical
  properties are completely given by its first two moments. It can be
  easily shown that
  $\langle \Delta F(t)\Delta
  F(t')\rangle=N^{-1}\gamma\,\delta(t-t')$,
  its variance tends to zero as $N^{-1}$, which is the typical
  behavior of fluctuating quantities in statistical mechanics. Even
  so, it should be noted that in biomolecules $N$ is not necessarily
  very large and certainly not of the order of Avogadro's number, and
  thus fluctuations play a major role.  In force--extension
experiments, the length is usually uniformly increased/decreased with
time $t$, $dL/dt=\mu$ with a constant $\mu$.

It is convenient to render our equations dimensionless. We set  the length unit $[\eta]$ equal to the difference between the extensions of the two free energy minima of a single unit for a certain applied force. It is natural to adopt the critical force, at which
the two minima are equally deep, as the unit of force,
$[F]=F_c$. The parameters $[\eta]$ and $[F]$ depend on the specific
choice of the double-well potential $a(\eta;Y,0)$. The 
free energy unit is then $[F]\,[\eta]$. We select the time scale
as $[t]=\gamma[\eta]/[F]$, where $\gamma$ is the typical friction
coefficient experienced by the units.  The typical value of
$\gamma$ can be obtained from the value of the diffusion coefficient
$D=T/\gamma$ of a single module protein being stretched  \cite{ber12}. In
principle, we introduce a new notation for the dimensionless
variables, $F^*=F/[F]$, etc.~but, in order not to clutter our
formulas, we drop the asterisks in the remainder of the paper.

\subsection{Fokker-Planck equation and equilibrium distributions}\label{F-P}

In force controlled experiments, $F(t)$ is a given function of time,
and the set of Langevin equations (\ref{1}) is equivalent to the
following Fokker-Planck equation for the probability density
$\calP(\bm{\eta},t)$ of finding the system with extension values
$\bm{\eta}=\{\eta_1,\ldots,\eta_N\}$ at time $t$,
\begin{equation}\label{1.4}
  \frac{\partial}{\partial t}\calP=\sum_{j=1}^N \frac{1}{\gamma_j} \frac{\partial}{\partial \eta_j} \left[\frac{\partial\calG}{\partial \eta_j} \calP \right]+  T \sum_{j=1}^N \frac{1}{\gamma_j} \frac{\partial^2 \calP}{\partial \eta_j^2}.
\end{equation}
where $\calG=\calA-F\calL$, as given by Eq.~\eqref{1.2a}.  If
the force $F$ is kept constant, Eq. (\ref{1.4}) has a stationary
solution, which is the statistical mechanics prescription,
\begin{equation}\label{1.5}
  \calP^{\eq}(\bm{\eta})\propto e^{- \calG(\bm{\eta};Y,F)/T }.
\end{equation}
Therefore, the equilibrium values of the module extensions $\bm{\eta}^{\eq}$ are the functions of $F$ that maximize $\calP$ or, equivalently, minimize $\calG$, that is, they verify
\begin{equation}\label{1.2c}
\left.  \left(\frac{\partial\calG}{\partial \eta_j}\right)_{\!\! Y,F} \right|_{\eq}=0 \,\, \Rightarrow \,\, \eta_j=\eta_j^{\eq}(Y,F), \quad j=1,\ldots,N.
\end{equation}
If there is only one minimum, this is the equilibrium configuration. If there is more than one, the absolute minimum is the thermodynamically stable configuration, while the other minima correspond to metastable states in the thermodynamic sense. For each equilibrium configuration, either stable or metastable,  the equilibrium value of the free energy $\calG$ is
\begin{equation}\label{1.2d}
  \calG^{\eq}(Y,F)=\calG(\bm{\eta}^{\eq}(Y,F);Y,F).  \
\end{equation}
Taking into account Eq.~\eqref{1.2c}, we have
\begin{equation}\label{1.2e}
  \left(\frac{\partial\calG^{\eq}}{\partial F}\right)_{\!\! Y}\!\!\!=
  \!\!\ 
  \left.\left(\frac{\partial\calG}{\partial
        F}\right)_{\bm{\eta},Y}\right|_{\eq} \!\!\!\! = \!
  -\sum_{j=1}^N \eta_j^{\eq}(Y,F)= \! -\calL^{\eq}(Y,F),
\end{equation}
which gives the equilibrium FEC under force control.

Let us consider now the length control situation. In the experiments,
the device controlling the length of the system does not have an
infinite stiffness and thus the length-control is not perfect, as
discussed above (see also \cite{PKTyS03} and Appendix~\ref{app:Berko}). {Had we taken} into account
this finite value of the stiffness $ \chi_{\text{lc}}$, the
Fokker-Planck equation {would have been} obtained by substituting the
Gibbs free energy $G\equiv A+\Delta U_{\text{fc}}$ in
Eq.~\eqref{1.4} by the corresponding thermodynamic potential
$A+\Delta U_{\text{lc}}$. Thus, the stationary solution of this
Fokker-Planck equation would be the equilibrium distribution
$\calP^{\eq}(\bm{\eta};Y,L)\propto \exp[- (\calA(\bm{\eta};Y)+\Delta
U_{lc}(\bm{\eta};L))/T]$.
Of course, in the limit as $ \chi_{\text{lc}}\to\infty$, the variance
of the Gaussian factor $\exp[-\Delta U_{lc}(\bm{\eta};L)/T]$ vanishes
and this factor tends to a delta function $\delta(L(\bm{\eta})-L)$
giving perfect length control.

In the case of perfect length control, the correct
Fokker-Planck equation {can} be obtained by taking the limit
  as $\chi_{\text{lc}}\to\infty$, but here we follow an alternative
  route. We calculate the first two moments of the extensions
$\bm{\eta}$, taking into account that not all the extensions $\eta_j$
are independent and that the force $F$ is given by eq.  (\ref{1.3c}),
\begin{eqnarray}
  \frac{\partial}{\partial t}\calP & = & \sum_{j=1}^N
  \frac{1}{\gamma_{j}}\frac{\partial}{\partial\eta_j} \left[ \left(
      \frac{\partial\calA}{\partial\eta_j}-F_{\text{FP}} \right)\calP
  \right] \nonumber \\
  && + T \sum_{j=1}^N \frac{1}{\gamma_{j}}\sum_{k=1}^N \left(
     \delta_{jk}-\frac{\gamma}{N\gamma_k} \right)
   \frac{\partial^2}{\partial\eta_j\partial\eta_k} \calP . \label{1.6}
\end{eqnarray}
Here $F_{\text{FP}}$ is given by Eq.~\eqref{eq:Fexp}.  If the length
is kept constant, $dL/dt=0$, Eq. (\ref{1.6}) has a stationary
solution,
\begin{equation}\label{1.7}
  \calP^{\eq}(\bm{\eta};Y,L)\propto \delta(L(\bm{\eta})-L) \; 
  e^{- \calA(\bm{\eta};Y)/T}
  ,
\end{equation}
as can be easily verified by inserting (\ref{1.7}) into
(\ref{1.6}). This means that $\calA$ is the relevant potential for the
statistical mechanics description at equilibrium, as was
expected. Eq.~\eqref{1.7} is consistent with the limit as
  $\chi_{\text{lc}}\to\infty$ of the equilibrium distribution for
  realistic length control, as already discussed above.

To obtain the equilibrium values for the extensions, we look
for the minima of $\calA$ with the constraint given by the delta
function in (\ref{1.7}), $\calL(\bm{\eta})=L$.  We have to introduce a
Lagrange multiplier $F$ and look for the minima of $\calA-F\calL$,
that is, the same minimization as in the force-controlled
case. However, the Lagrange multiplier is an unknown that must be
calculated at the end of the process by imposing the constraint,
$F=F(L)$. This Lagrange multiplier is, from a physical point of view,
the force that must be applied to the system in order to have the
desired length.  The equilibrium extensions $\eta_j^\eq(L)$ are thus
given by the solutions of
\begin{equation}\label{1.2fb}
  \left. \left(\frac{\partial\calA}{\partial \eta_j}\right)_Y\! \right|_\eq\!\!=F, \,\, j=1,\ldots,N; \,\, \sum_{j=1}^N \eta_j^\eq(Y,F)=L.
\end{equation}
The last equation gives the FEC, $L=L(Y,F)$ or
$F=F(Y,L)$, from which we obtain
$\eta^\eq=\eta^\eq(Y,L)$. The thermodynamic potential $\calA^{\eq}$ is
the Legendre transform of $\calG^{\eq}$ with respect to $F$. In fact,
the equilibrium value of $\calA$, $\calA^{\eq}(Y,L)=\calA(\bm{\eta}^{\eq}(Y,L);Y)$,
verifies that
\begin{equation}\label{1.2g}
\left(  \frac{\partial\calA^{\eq}(Y,L)}{\partial L}\right)_{Y}=
F.
\end{equation}
The proper variables for $\calA^{\eq}$ are the set of intensive
parameters $Y$ (temperature $T$, pressure $p$, $\ldots$ of the fluid
in which the polyprotein is immersed) and the extensive length $L$
\cite{footnote1}, while the proper variables for $\calG^{\eq}$ are all
intensive, $Y$ and $F$. In this sense, $\calA^{\eq}$ plays the role of
Helmhotz free energy, while $\calG^{\eq}$ is the analogous of Gibbs
free energy. It should be stressed that (i) however, different
notations are found in the literature for these two thermodynamic
potentials; (ii) as in the case of magnetic systems \cite{tho88},
there is a difference of sign with respect to the usual free energy
terms with the pressure $p$ and the volume $V$.

{The f}luctuation theorems for Markov processes described by the
  Langevin (or the equivalent Fokker-Planck) equations have been
  thoroughly analyzed in Ref.~\cite{CCyJ06}. The results therein are
  directly applicable to the Fokker-Planck equations derived here for
  the force-controlled and the realistic (finite $\chi_{\text{lc}}$)
  length-controlled cases. Whe{n the controlled parameter (either force or length)} is kept constant (time-independent),
  detailed balance applies and the corresponding stationary
  distributions are equilibrium (canonical) ones {(in the
  terminology of section 2 in Ref.~\cite{CCyJ06})}. In the limit
as  $\chi_{\text{lc}}\to\infty$, we expect this result to be still valid
  on physical grounds, but further mathematical work would be
  necessary to establish it rigorously: Some of the matrices defined
  in \cite{CCyJ06} become
  singular and thus have no inverse. This is a point that certainly
  deserves further investigation, but it is out of the
  scope of the present paper.

\section{The ideal chain}\label{ideal_chain}

In this Section, we analyze the case of an ideal chain, in which the
identical units do not interact either among themselves or
with the cantilever/platform, $k_j=0$ and $\delta_j=0$ for all $j$
\cite{pra13}. We analyze the equilibrium situation and thus solve the
minimization problems for the force-controlled and
length-controlled cases of the previous section. We
also investigate the dynamical situation arising in processes
  in which the force or length varies in time at a finite rate, and
compare these dynamical FECs to the equilibrium
ones.

\subsection{Double-well potential. Equilibrium branches.}
\label{landau_free_energy}

In order to keep the notation simple, we omit the dependence on the
intensive parameters $Y$ of the free energy parameters. As a
  minimal model, we consider
the polynomial form, \`a la Landau, for the free energy
\cite{pra13}
\begin{equation}\label{fe1}
   \calA(\bm{\eta})=\sum_{j=1}^N a(\eta_j), \quad a(\eta)=F_c \eta - \alpha \eta^2+\beta \eta^4.
\end{equation}
The parameters $F_c$, $\alpha$ and $\beta$ are all positive functions
of the intensive parameters $Y$. Specifically, $F_{c}$ plays
  the role of the critical force, above (below) which the unfolded
  (folded) configuration is the most stable one, as shown in what
  follows. The possible equilibrium extensions $\eta^\eq$ are the
minima of $a(\eta)-F\eta$,
\begin{equation}\label{fe1b}
  a'(\eta^{(i)})=F
\end{equation}
or, equivalently,
\begin{equation}\label{fe2}
  -2 \alpha \eta^{(i)} +4 \beta \left(\eta^{(i)}\right)^3=\varphi, \quad \varphi\equiv F-F_c.
\end{equation}
We have introduced the notation $\ei$ because Eq. (\ref{fe2}) has
three solutions in the metastability region, given by
$|\varphi|=|F-F_c|< \varphi_0=(2\alpha/3)^{3/2} \beta^{-1/2}$.  We set
the indexes by choosing $\1<\2<\3$. They depend on the force $F$
through $\varphi$ (and on the intensive variables $Y$ through
$\{\alpha,\beta,F_c\}$).  The extensions $\1(\varphi)$ and
$\3(\varphi)$ are locally stable because they correspond to minima of
$a_j-F\eta_j$, while $\2(\varphi)$ corresponds to a maximum and is
therefore unstable.  The curvatures at the folded and unfolded states
are
$\chi^{(i)}(\varphi)=a''(\ei(\varphi))=12\beta[\ei(\varphi)]^2-2\alpha$,
$i=1,3$.  Both curvatures (i) are positive in the metastability region
and (ii) vanish at their limits of stability, $\chi^{(1)}$
($\chi^{(3)}$) at $\varphi=\varphi_0$ ($\varphi=-\varphi_0$).

The situation is similar to that analyzed by Landau \cite{landau} for
a second order phase transition under an external field, with $\eta$
and $\varphi=F-F_c$ playing the role of the order parameter and the
external field, respectively. At the critical force $\varphi=0$, the
stable equilibrium values of the extensions are
\begin{equation}\label{fe7}
  \3_c=-\1_c=\left( \frac{\alpha}{2\beta} \right)^{1/2}.
\end{equation}
They are equiprobable, since $g(\eta)=a-F\eta$ is an even function of $\eta$
for $F=F_{c}$, and $\guno=\auno-F_c\1=\gtres=\atres-F_c\3$, where we have
introduced the notation $\auno\equiv a(\1)$, $\atres\equiv
  a(\3)$, $\guno\equiv g(\1)$, and $\gtres\equiv g(\3)$.
For $F\neq F_c$, the ``field'' $\varphi$ favors the state with
$\sgn(\varphi)=\sgn(\eta)$. In fact, at the limit of
  stability we have that $\guno=-13\alpha^{2}(6\beta)^{-1}$ for
  $\varphi=-\varphi_{0}$ (or $\gtres=-13\alpha^{2}(6\beta)^{-1}$ for
  $\varphi=\varphi_{0}$). Therefore, in the metastability region
$|\varphi|<\varphi_0$, we have the following picture: For $F<F_c$, the
thermodynamically stable state is the folded one $\1<0$ and the
unfolded one $\3>0$ is metastable. For $F>F_c$, the situation is
simply reversed.  On the other hand, the folded $\1$ (unfolded $\3$)
state also exists for forces below (above) the metastability region
$\varphi<-\varphi_0$ ($\varphi>\varphi_0$). In their respective
regions of existence, both locally stable extensions $\1$ and $\3$ are
increasing functions of $\varphi$ (or $F$), since Eq.~\eqref{fe1b}
implies that $\chi^{(k)}(\varphi)d\eta^{(k)}/d\varphi=1$.  At zero
force, one module can be folded or unfolded if
$\varphi_0>F_c$, while we have only the folded state if
$\varphi_0<F_c$.

\begin{figure}
\begin{center}
    \includegraphics[width=3.25in]{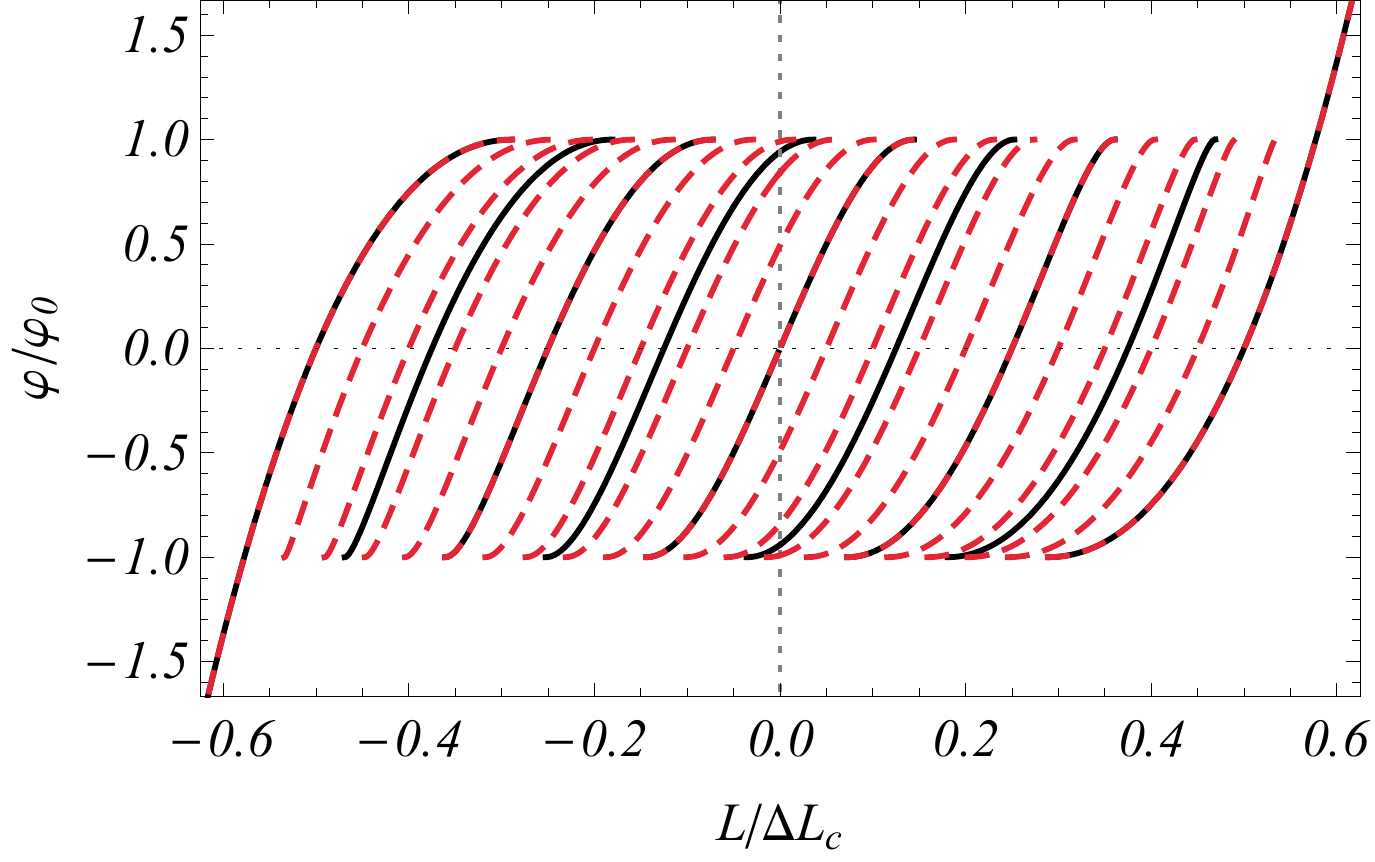}
    \caption{Normalized FECs for $N=8$ (solid black) and $N=20$
      (dashed red).  Zero length corresponds to having half of the
      units unfolded at $F=F_c$. There are $N+1$ branches in the
      metastability region $|\varphi/\varphi_0|< 1$, with the number
      of unfolded units $J$ increasing from left to right.  The first
      ($J=0$) and last ($J=N$) branches are independent of $N$.  Note
      that the branches become denser as $N$ increases, {and
        also the up-down and left-right symmetry thereof. These
        symmetries stem from the simple form of the Landau-like free
        energy \eqref{fe1}, and thus they are not present if a more
        realistic potential is considered, see Appendix \ref{app:Berko}.} }
    \label{ramas_eq}
\end{center}
\end{figure}

Either module can be either folded  or unfolded in the metastability region, and thus a FEC with $N+1$ branches shows up, as seen in Fig.~\ref{ramas_eq}.  The $J$-th branch of the $F-L$ curve corresponds to $J$ unfolded modules and $N-J$ folded ones, $J=0,\ldots,N$. Since there is no coupling among the units, the equilibrium value of $\calA$ 
over the $J$-th branch is
\begin{subequations}\label{fe8}
\begin{equation}\label{fe8a}
  \calA_J^\eq=(N-J)\auno +J \atres .
\end{equation}
The corresponding length is
\begin{equation}\label{fe8b}
  \calL_J = (N-J) \1+J\3\!.
\end{equation}
\end{subequations}
Both $\calA_J^\eq$ and $\calL_J$ are functions of $F$ and the
intensive parameters $Y$ through the equilibrium
extensions. Eq.~\eqref{fe8b} is the FEC, both for the force and length
controlled cases. In Fig.~\ref{ramas_eq}, we have normalized
  the lengths with
\begin{equation}\label{fe11}
  \Delta \calL_c=\calL_N(F_{c})-\calL_0(F_{c})=N \left(\3_c-\1_c\right),
\end{equation}
which is the difference of lengths between the completely
  unfolded branch ($J=N$) and the completely folded one ($J=0$) at the
  critical force. It is interesting to note that similar multistable
equilibrium curves appear in quite different physical systems: from
storage systems \cite{dre10,dre11,dre11cmt} to semiconductor
superlattices \cite{GHMyP91,RTGyP02,BGr05,BT10,BHGyT12}. For instance,
see Fig.~3 of Ref.~\cite{dre10} and Fig.~6 of Ref.~\cite{dre11} for
the chemical potential vs.~charge curve in storage systems, and
Fig.~8.13 of Ref.~\cite{BT10} for the current-voltage curve of a
superlattice.

As discussed in the previous section, we have chosen $[F]=F_c=1$ and
$[\eta]=\3_c-\1_c=1$ as units of force and length. Using
Eq.~\eqref{fe7}, $\beta=2\alpha$ and $\3_c=-\1_c=1/2$.  Moreover, the
folded state $\1$ is the most stable one at zero
force. This means that the unstable state $\2$ is closer to
the metastable state $\3$ for the simple Landau potential we are using
\cite{note}. For the sake of concreteness, we take $\2-\1=0.9(\3-\1)$
at zero force, which leads to $\alpha=273^{3/2}/1672\approx 2.697787$
and $\varphi_0=91^{3/2}/836=1.038378$. It should be stressed that all
the \textit{normalized} plots in this section are independent of this
particular choice of parameters. A more conventional definition of
protein length could be to select at zero force (a) zero extension for
the folded modules (b) the difference between the unfolded and folded
configurations as the length unit. This `physical' definition would
give a nondimensional extension
\begin{subequations}
  \begin{equation}
    \label{eq:u_zero_force}
    u=\frac{\eta-\1(F=0)}{\3(F=0)-\1(F=0)},
  \end{equation}
  and a polyprotein length
\begin{equation}
L_u=\frac{-N\1(F=0)+\sum_{j=1}^N\eta_j}{\3(F=0)-\1(F=0)}\!\!=\frac{19N}{30}+\frac{\sqrt{273}}{15} L,\label{phys_length}
\end{equation}
\end{subequations}
respectively. At zero force, the module extensions are
$u^{(1)}=0$, $u^{(2)}=0.9$ and $u^{(3)}=1$. The length $L_u$ is
typically positive for $F>0$, that is, for $\varphi>-F_{c}$.

We expect that the simple Landau-like free energy given by
Eq.~\eqref{fe1} should be relevant to investigate qualitatively the
FECs for forces/lengths close to the metastability region. In
particular, this minimal choice does not account for the existence of
a maximum length of the polymer, its so-called contour length
\cite{length}, a fact that becomes significant for high forces. In
order to study the whole range of forces and/or try to describe
quantitatively the experiments, we should use a more realistic
potential, such as that proposed by Berkovich et al.~\cite{ber10} or
modifications thereof. We did this in Ref.~\cite{BCyP14} to understand
the stepwise unfolding observed in force-clamp experiments. The simpler potential used in this paper suffices for: 
(i) showing that the key aspects
of the experimental behavior observed in the unfolding/refolding
region can be understood within a minimal model, and (ii) establishing
connections with other physical systems such as storage devices
\cite{dre10,dre11,dre11cmt} or semiconductor superlattices
\cite{GHMyP91,RTGyP02,BGr05,BT10,BHGyT12} that have similar behavior
in the metastability region. We briefly investigate an asymmetric potential in Appendix \ref{app:Berko} to understand why the experimentally observed FEC corresponding to unfolding under length-control is reproduced by the Landau-like potential whereas the FEC corresponding to refolding is not, see Sec.~\ref{cont_length_eq} for details.

\subsection{Force control}\label{cont_force_eq}

In force-controlled experiments, the Gibbs free energy is the relevant
thermodynamic potential because it appears in the equilibrium
distribution (\ref{1.5}). As discussed in Sec.\/ \ref{F-P}, the stable
state corresponds to the absolute minimum of $\calG$. All the units in
our ideal chain are independent under force control. Therefore, by
increasing quasi-statically the force, the equilibrium FEC
(\ref{fe8b}) is swept.  Over the $J$-th branch with $J$ unfolded
modules,
\begin{equation}\label{fe9}
  \calG^\eq_J=
  (N-J) \guno + J \gtres, \quad \gi=\ai-F\ei.
\end{equation} 
For $F<F_c=1$ ($F>F_{c}$), the absolute minimum of
$\calG$ corresponds to the folded (unfolded) state $\1$
($\3$) and the system moves over the force--extension branch
in which none (all) of the units are unfolded, $J=0$
($J=N$).

\begin{figure}[htbp]
    \begin{center}
    \includegraphics[width=3in]{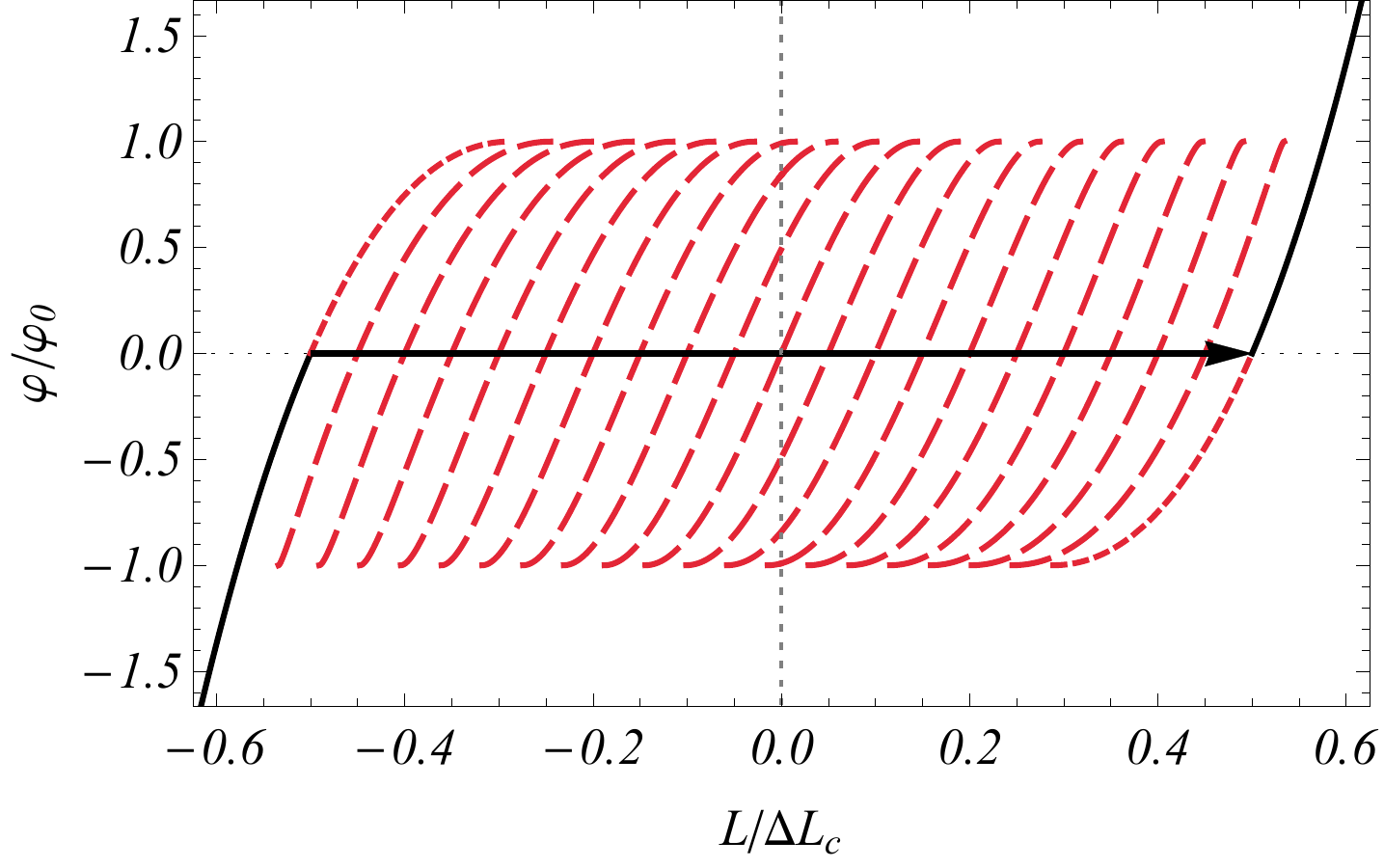}
    \includegraphics[width=3in]{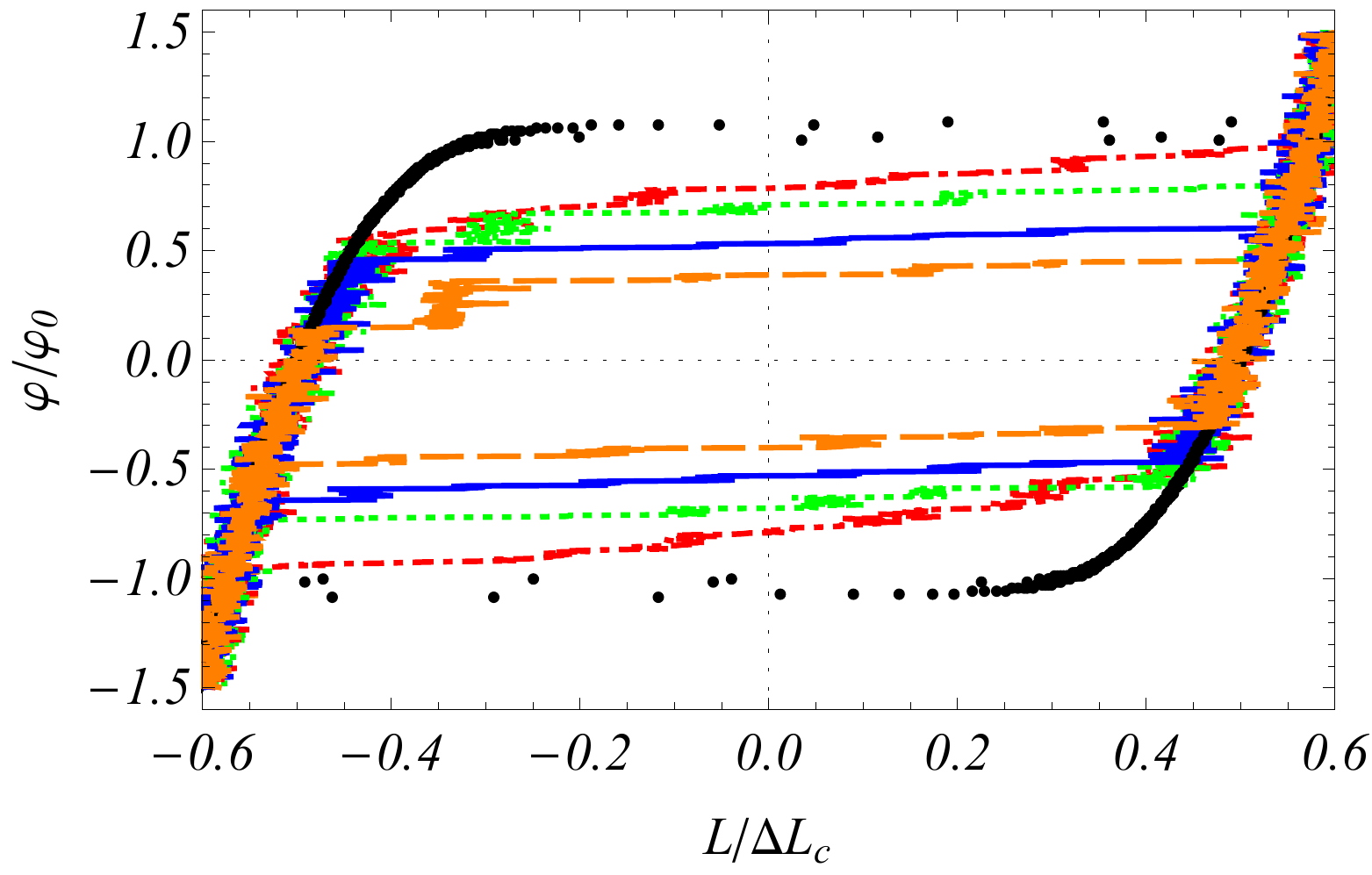} 
    \caption{(Top) First order transition in the length for a
      quasi-static increase of the force.  We use different colors for
      the stable parts of the branches (solid black) and the
      metastable parts (dashed red). The first branch $J=0$ is swept
      until the critical force $\varphi=0$ is reached. Then all the
      modules unfold simultaneously and the system goes directly to
      the completely unfolded branch $J=N=20$ (arrow).  (Bottom)
      Hysteresis cycles under force-controlled conditions for a $N=20$
      system. The lines correspond to simulations of the
        Langevin equations \eqref{langevin_eqs} for the temperature
      $T=0.02$ and different rates of variation of the force, namely
      $|dF/dt|=3\times 10^{-k}$, with $k=2$ (dot-dashed red), $k=3$
      (dotted green), $k=4$ (solid blue), and $k=5$ (dashed
      orange). The same rates of variation of the force are
        considered for the very low temperature $T=2\times
      10^{-5}$. All the curves are superimposed and thus they are
        plotted with the same symbols (black dots).  For the higher
      temperature, the area of the hysteresis cycle decreases with the
      rate, approaching the behavior for a quasi-static process.  }
    \label{disc_L}
    \end{center}
\end{figure}

Unfolding is a first-order phase transition between these states that
occurs at the critical force $F_c=1$ defined by continuity of forces
and of the Gibbs free energies,
$\calG^\eq_0|_{F_c}=\calG^\eq_N|_{F_c}$.  At $F_c=1$, all the units
unfold simultaneously. The length, which is a function of $F$ given by
Eq.~\eqref{1.2e}, has a discrete jump equal to $\Delta L_{c}$,
  given by Eq.~\eqref{fe11}.  It is worth recalling $\3_c-\1_c=1$ in
nondimensional units. The free energy (\ref{fe9}) produces
\begin{equation}\label{fe12}
  \frac{d}{dF} \left( \calG_N^\eq-\calG_0^\eq \right)=- N \left(\3(F)-\1(F)\right)<0, \quad \forall F,
\end{equation}
consistently with Eq.~\eqref{1.2e}. Then the basin of attraction of
the completely folded branch is the largest one for $F<F_c$, whereas
the completely unfolded branch has the largest basin of attraction for
$F>F_c$.  All the intermediate metastable branches with $J\neq 0,N$
are not ``seen'' by the system in a quasi-static process that takes
infinite time to occur, see the top panel of
Fig. \ref{disc_L}.

For a real, non-quasi-static process, the simple equilibrium picture
above is not realized.  Depending of the rate of variation of the
force and the strength of the thermal fluctuations, the system will
explore the metastable branches of the FEC. Then intermediate states
between the completely folded and unfolded configurations will be seen
\cite{hug10,rit06jpcm,Ka12,pra12}. This is shown in the bottom panel
of Fig.~\ref{disc_L} by solving (\ref{1})-(\ref{1.3}) with
$k_j=\delta_j=0$ (in nondimensional form) for a 20-module protein,
with $T=2\times 10^{-5}$ and $T=0.02$. All the $T=2\times 10^{-5}$
curves are superimposed on each other because the considered rates
$|dF/dt|$ are small enough to lead to the \textit{adiabatic} limit.
For the upsweeping (downsweeping) process, the system moves over the
completely folded (unfolded) branch until it reaches the end thereof,
$\varphi=\varphi_0$ ($\varphi=-\varphi_0$). Then it jumps to the
completely unfolded (folded) branch. The temperature is so small that
the activated processes over the free energy barriers take place over
a much longer time scale. For the higher temperature, $T=0.02$, the
system can jump between the different minima of the potential and the
force at which the system jumps between branches depends on the rate
of variation of the force. Also, the system partially explores some of
the intermediate branches.  This picture is consistent: quite close to
the adiabatic limit, the hysteresis cycle is large for the highest
rate of variation, whereas the cycle shrinks towards the straight line
$\varphi=0$ ($F_c=1$) as the rate tends to zero.

\subsection{Length control}\label{cont_length_eq}

\begin{widetext}

\begin{figure}[htbp]
\begin{center}
    \includegraphics[width=3in]{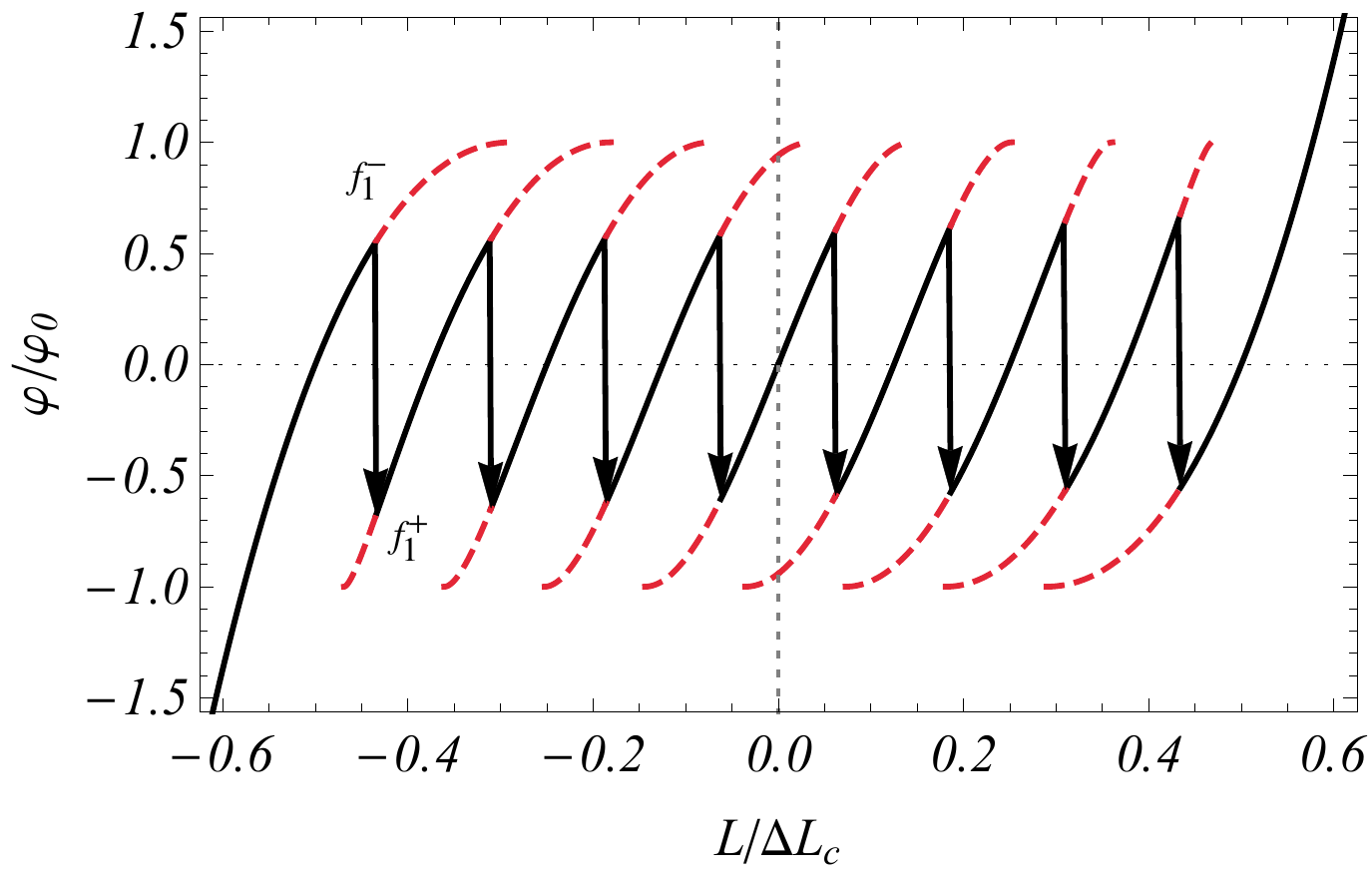}
    \includegraphics[width=3in]{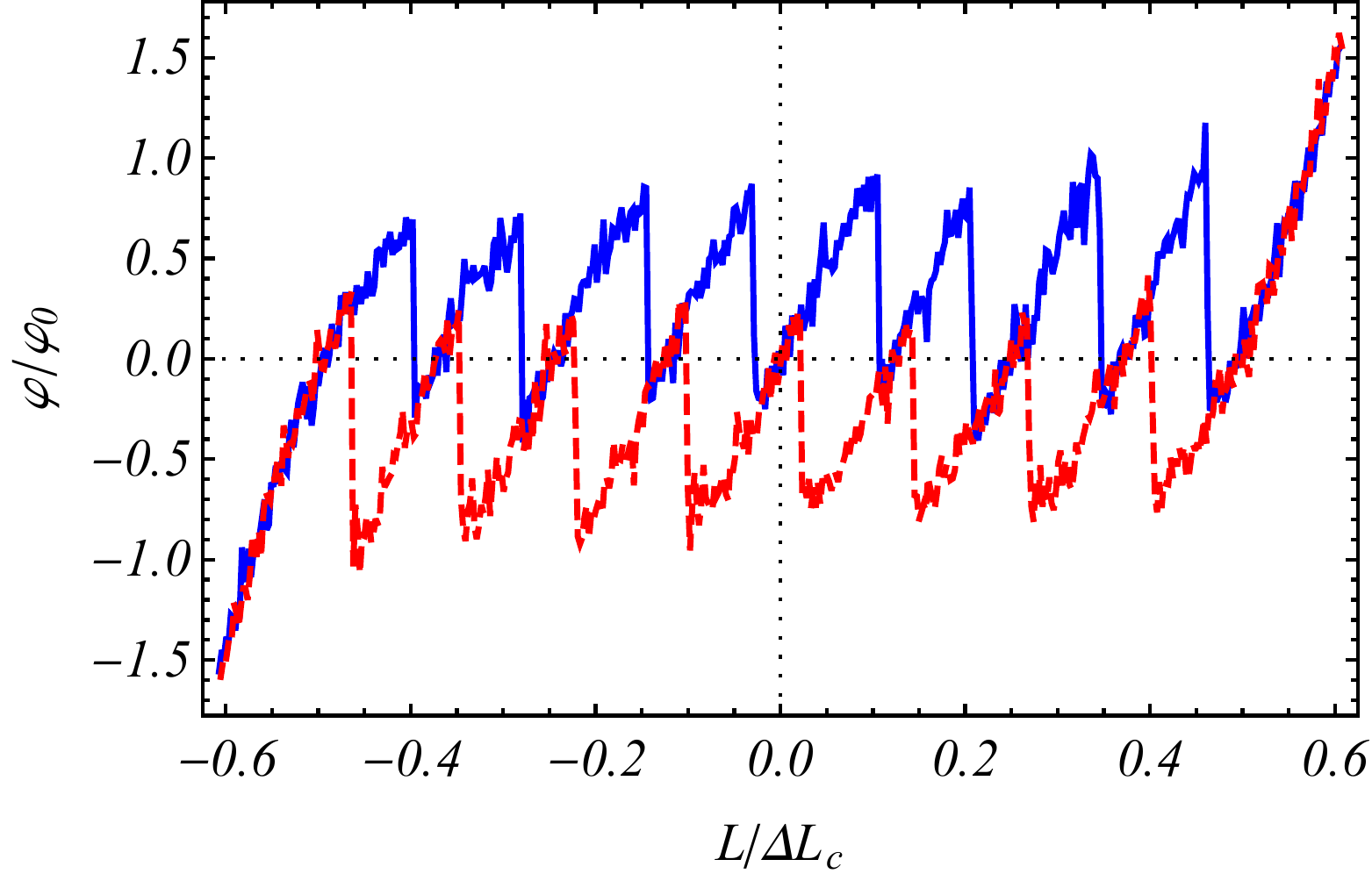}
    \includegraphics[width=3in]{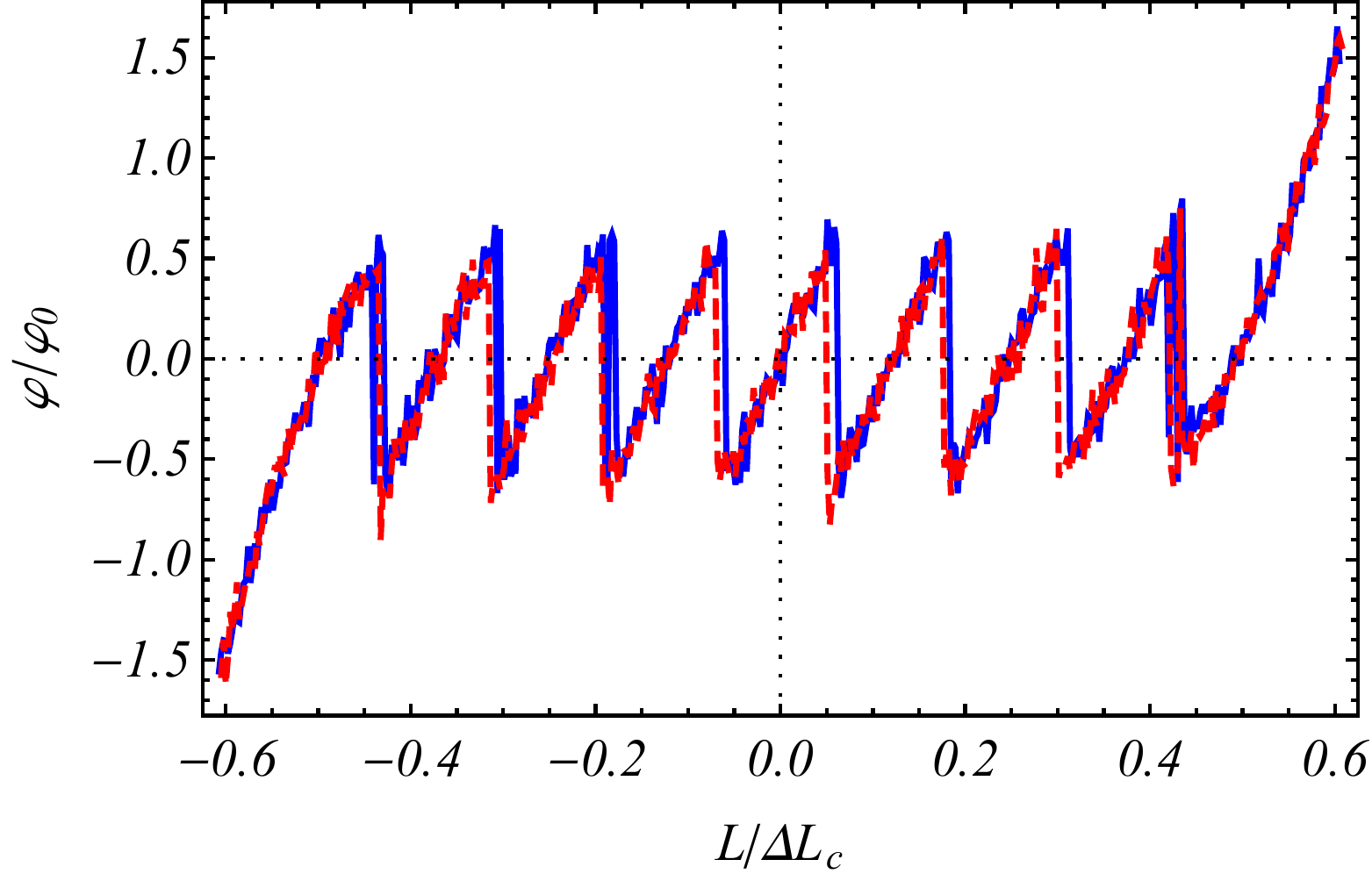}
    \includegraphics[width=3in]{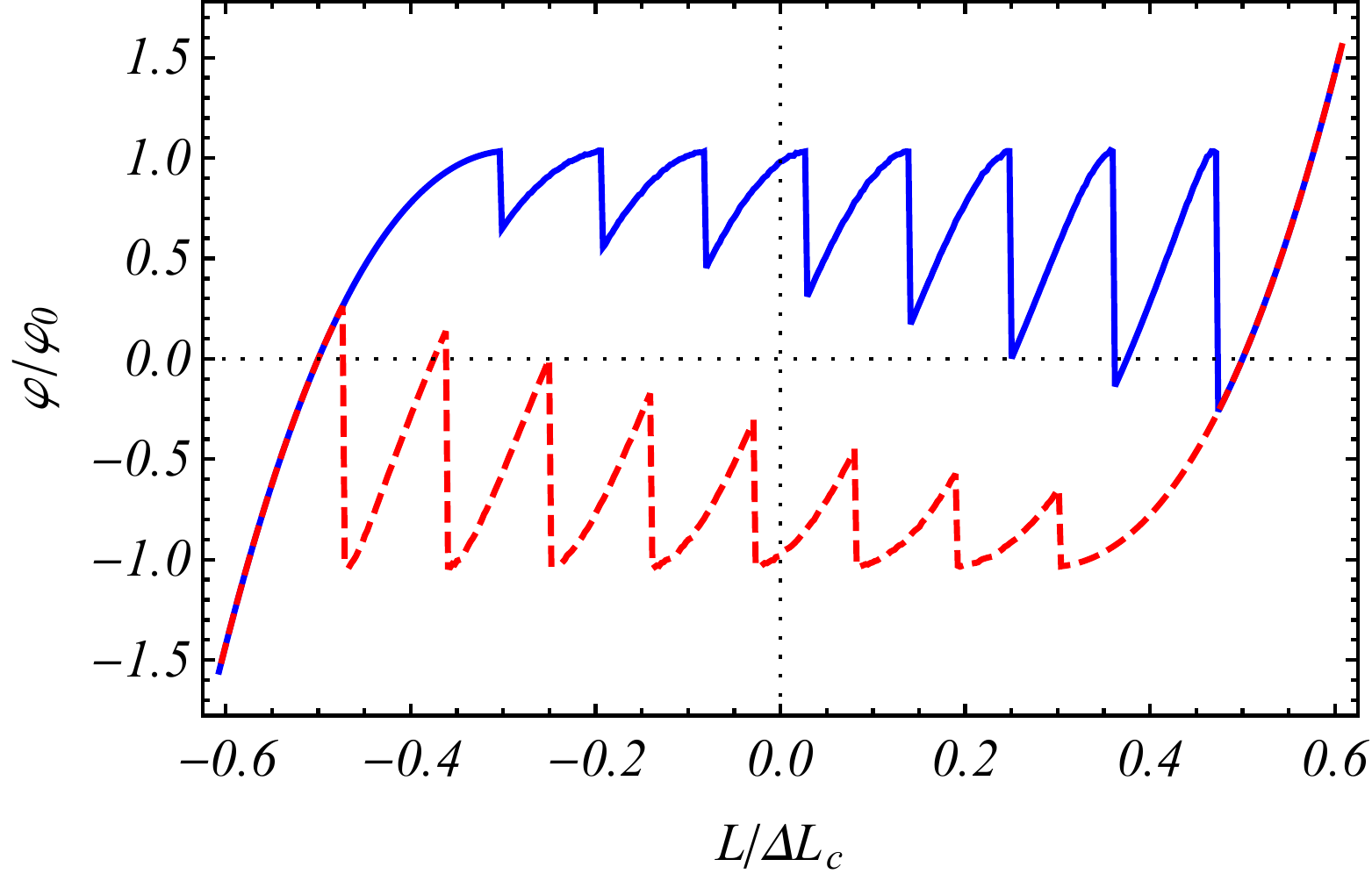}
    \caption{(Top left) Equilibrium force rips in the $F-L$ curve for
      a system with $N=8$ domains. We use different colors for the
      stable parts of the branches (solid black), metastable parts
      (dotted red), and the force rips (black arrows).  The system
      follows the solid black curve in a quasi-static pulling process,
      with a series of first order transitions in the force (marked by
      the arrows). At the $J$-th transition, the force changes from
      $f_J^-$ (over the $(J-1)$th branch) to $f_J^+$ (over the $J$th
      branch). These forces $f_J^{\pm}$ increase with the number of
      unfolded units, as observed in AFM experiments with modular
      proteins, even though all the units are perfectly identical in
      the model. (Top right) Hysteresis cycle for a system composed of
      $N=8$ modules. The dimensionless temperature $T=0.02$, and the
      rate of variation of the length is $|dL/dt|=1.2\times 10^{-3}$
      ($\dot{L}>0$, solid blue; $\dot{L}<0$, dashed red). (Bottom
      left) The same as in the top right panel, but for a smaller rate
      $|dL/dt|=1.2\times 10^{-6}$. Aside from thermal fluctuations,
      the system almost sweeps the equilibrium curve. (Bottom right)
      The same plot as in the bottom left panel, but for
      $T=2\times 10^{-5}$. Thermal fluctuations are so small that the
      system approaches the $T=0$ behavior, in which the branches are
      swept up to the end of the metastability region. }
\label{force_rips}
\end{center}
\end{figure}

\end{widetext}

In length-controlled experiments, the length constraint introduces a
long-range interaction between the protein modules. The equilibrium
probability of any configuration $\bm{\eta}$ is now given by
Eq.~(\ref{1.7}). Then the equilibrium configuration
$\bm{\eta}^\eq$ is found by minimizing $\calA$ with the constraint
(\ref{1.2b}), and the difference between values of $\calA^\eq$ at
adjacent branches in the $F-L$ diagram governs the stability thereof.
The length $\ell_J$ at which there is a change in the relative
stability of two consecutive branches, with $J-1$ and $J$ unfolded
units, is determined by the equality of their respective free energies
$\calA^\eq$. The corresponding forces $f_J^-\equiv F_{J-1}(\ell_J)$
and $f_J^+=F_J(\ell_J)$ over the branches with $J-1$ and $J$ unfolded
units obey the system of two equations
\begin{equation}\label{fe13}
\left. \calA_{J-1}^\eq\right|_{f_J^-}\!=\left. \calA_{J}^\eq\right|_{f_J^+}, \quad
\left. \calL_{J-1}^\eq\right|_{f_J^-}\!=\left. \calL_{J}^\eq\right|_{f_J^+}.
\end{equation}
The force rips at $L=\ell_J$ are $N$ first-order equilibrium phase
transitions because (i) the thermodynamic potential $\calA^\eq$ is
continuous at the transition, (ii) $F=(\partial\calA^\eq/\partial
L)_Y$ has a finite jump, from $f_J^-$ to $f_J^+<f_J^-$ at the $J$-th
transition. In the top left panel of Fig.\ \ref{force_rips}, we
explicitly show $f_1^-$ and $f_1^+$. We have the following picture: As
observed in Fig.~\ref{ramas_eq}, the branches $J-1$ and $J$ coexist on
a certain range of lengths. Inside this range, Eq.~(\ref{1.2g})
implies
\begin{equation}\label{fe14}
  \left(\frac{\partial}{\partial L} \left[\calA^\eq_{J}-\calA^\eq_{J-1} \right]\right)_Y=F_J(L)-F_{J-1}(L)<0,
\end{equation}
where we have used (\ref{1.2fb}). At equal length values $L$, the
force is larger on the branch with a smaller number of folded units,
$F_J(L)<F_{J-1}(L)$, $\forall J$. Therefore,
$\calA^\eq_{J-1}<\calA^\eq_{J}$, and then the branch $J-1$ is the
stable one and $J$ is metastable for $L<\ell_J$. The situation
reverses for $L>\ell_J$, and there are not more stability changes
between these branches because $\calA^\eq_{J}-\calA^\eq_{J-1}$
decreases monotonically as a function of $L$, as given by
(\ref{fe14}). Each intermediate branch ($J=1,\ldots,N-1$) is thus
stable between $\ell_J$ and $\ell_{J+1}$, that is, between $f_J^+$ and
$f_{J+1}^-$ (see top left panel of Fig. \ref{force_rips}). A sawtooth pattern arises in
the $F-L$ curve, with $N$ transitions between the $N+1$ branches at
lengths $\ell_1,\ldots,\ell_{N}$.

Similarly to the analysis for the force-controlled case, let us
investigate the behavior of the system when the length is first
increased and afterwards decreased with the same rate.  Depending on
the rate and the value of the temperature, a region of the metastable
part of the branches is explored, and the force rips do not take place
at the equilibrium values $\ell_{J}$. In Fig.~\ref{force_rips}, apart
from the equilibrium force-extension curve (top left panel), we plot
three {unfolding/refolding} cycles for an ideal 8-module protein.  In
the top right panel and the bottom left panels, the temperature is
$T=0.02$ and the rates are $|dL/dt|=1.2\times 10^{-3}$ and
$1.2\times 10^{-6}$, respectively. For the smallest rate, the system
basically sweeps the equilibrium curve, aside from thermal
fluctuations. Note that some of the transitions are ``blurred''
because of the hopping between the two posible forces at the
transition length for this very small rate. On the other hand, for the
highest rate, some hysteresis is present. Finally, in the bottom right panel, the
temperature is much lower, $T=2\times 10^{-5}$, which results in the
largest hysteresis cycle. This low temperature dynamical FEC is
basically the same for the two rates considered before,
$|dL/dt|=1.2\times 10^{-3}$ and $|dL/dt|=1.2\times 10^{-6}$, so only
the latter is shown. The system sweeps each branch up to its limit of
(meta)stability, $|\varphi/\varphi_{0}|=\pm 1$.  Interestingly, this
low temperature behavior resembles that of the chemical potential in a
recent investigation of the thermodynamic origin of hysteresis in
insertion batteries \cite{dre10,dre11}. This indicates that thermal
fluctuations are less relevant for insertion batteries than for
modular proteins, despite the similarities in their mathematical
description. In our simulations, we have averaged the force over a
unit time interval to mimic the experimental situation, in which the
measuring devices have finite resolution \cite{footnote2}.

In the unfolding process, the transitions occur at forces/lengths that
are displaced upward with respect to that of the refolding process, as
usually observed in the experiments
\cite{rit06jpcm,lip01,hug10,OMCyF99,RPSyG99,SSHyR02,man05,lee10}. However,
the unfolding/refolding curves for the simple Landau-like quartic
potential we are using are much more symmetric than the experimental
ones, reflecting the symmetry of the potential. In the experiments
with modular proteins, the unfolding FEC exhibits large force
rips similar to ours, but the refolding FEC does not present
a sawtooth pattern \cite{OMCyF99,RPSyG99,SSHyR02}. Recently, however,
several force rips have been observed in the refolding of the NI6C
protein, both in AFM experiments and Steered Molecular Dynamics
simulations \cite{lee10}. In Appendix \ref{app:Berko}, we briefly
analyze the predictions of our theory for the more realistic potential
introduced in Ref.~\cite{ber10}. In this case, the unfolding and
refolding curves are strongly asymmetric and closely resemble the
experimental ones.
  
\begin{figure}[htbp]
\begin{center}
    \includegraphics[width=3in]{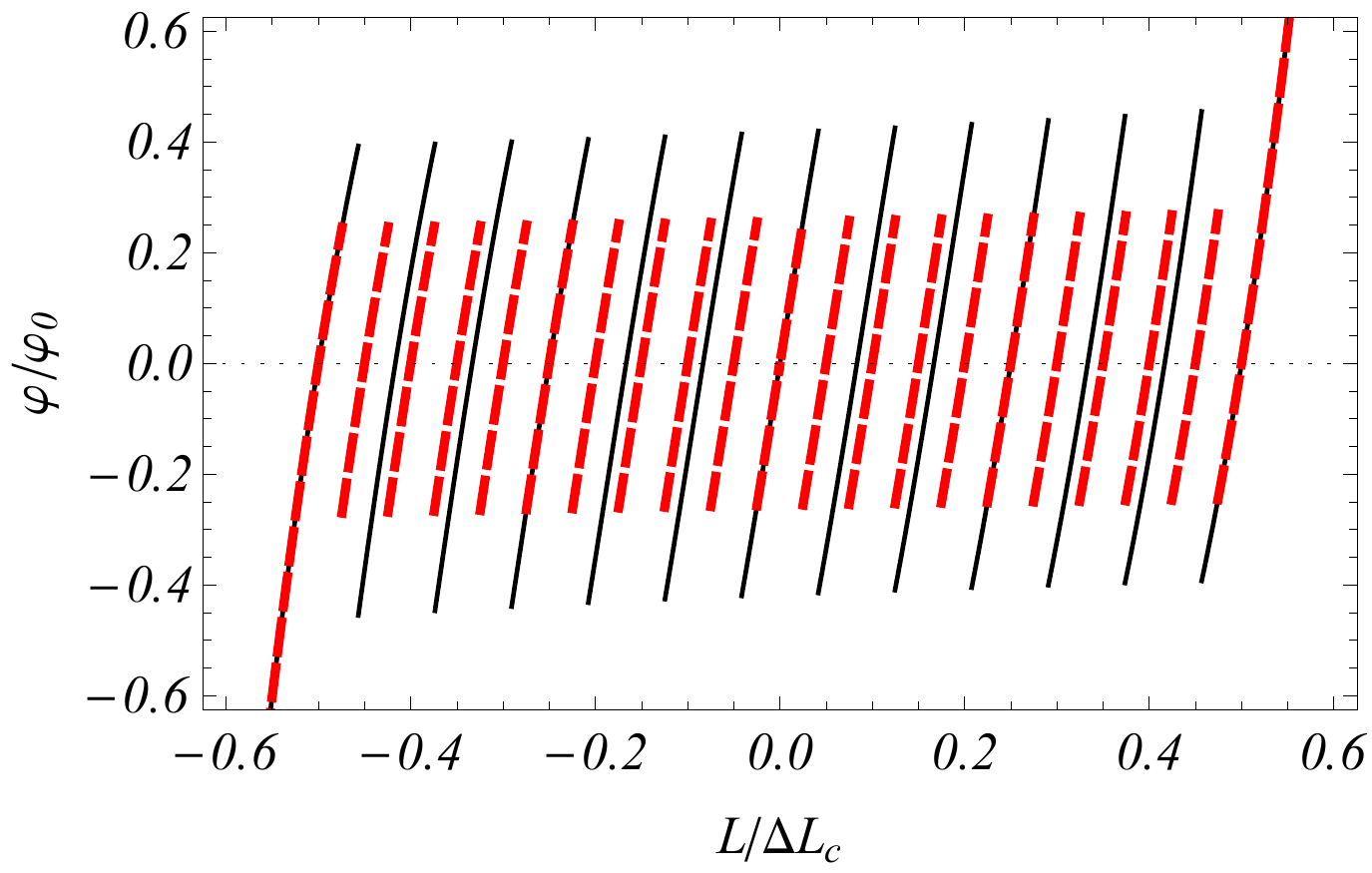}
\includegraphics[width=3in]{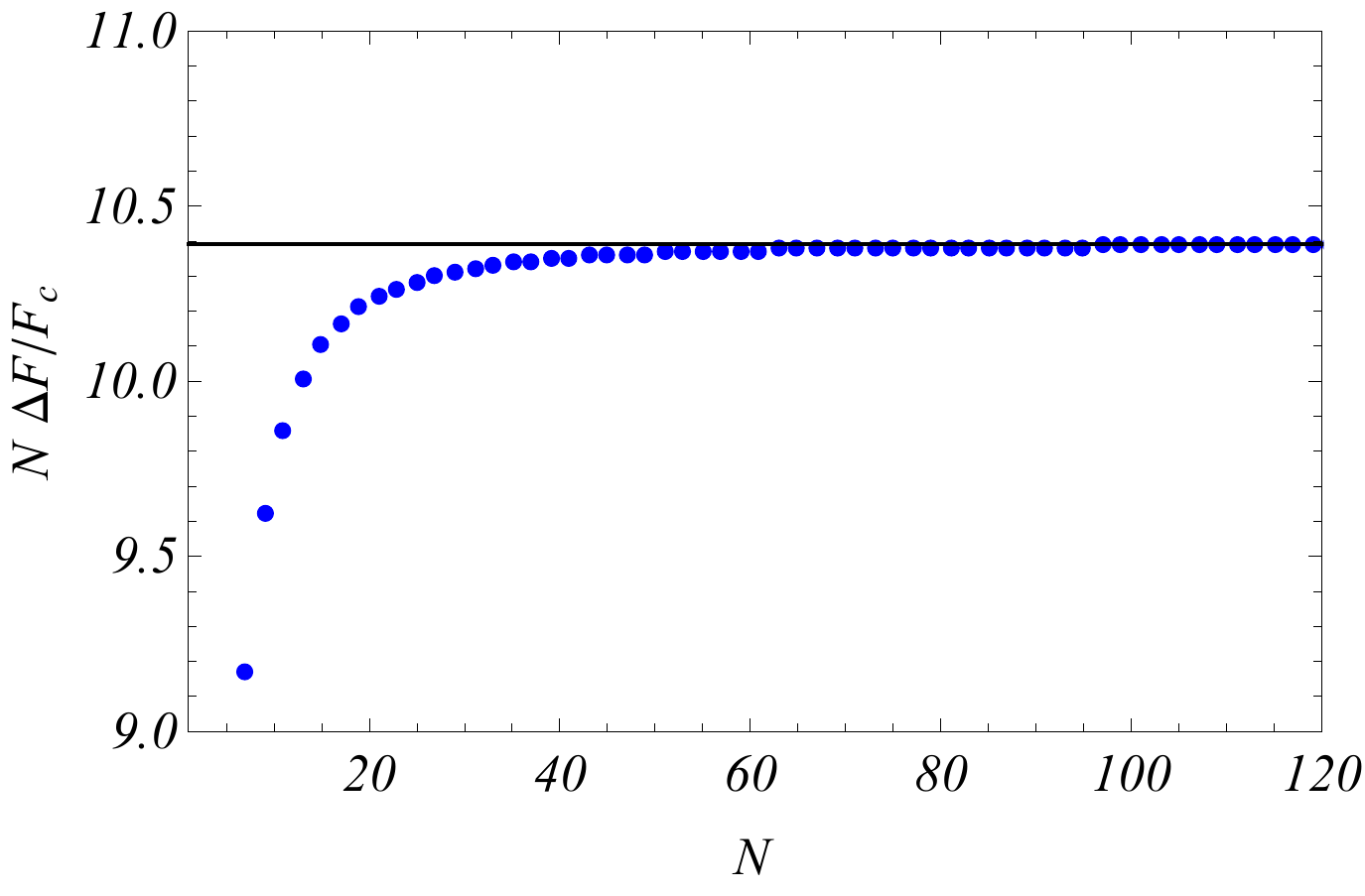}
\caption{(Top panel) Zoom of the metastability region for two chains
  with $N=8$ (solid line) and $N=20$ (dashed line). The size of the
  rips decreases as the number of units $N$ increases, and it vanishes
  as $N\to\infty$. (Bottom panel) Decrease of the force rips with the
  number of units $N$. We plot the size of the rips
  $\Delta F=f_{J}^{-}-f_{J}^{+}$ for the ``central'' transition with
  $J=(M+1)/2$, scaled with the factor $N/F_{c}$ (circles). The
  limiting value $6\sqrt{3}$, which represents the power law behavior
  given Eq.~\eqref{fe15}, is shown with a solid line. It is observed
  that the system approaches rapidly this asymptotic behavior, being
  very close to it for $N\gtrsim 20$.}
    \label{rips_size}
    \end{center}
\end{figure}

For both, equilibrium and dynamical FECs (the latter being closer to
the real experimental situation), (i) the size of the force rips
decrease with the number of units $N$, and (ii) $f_{J}^{\pm}$ increase
with the number of unfolded units $J$ for moderate values of $N$. The
equilibrium case is illustrated by the top panel of
Fig.~\ref{rips_size}. Interestingly, the increase with $J$ of the rips
forces has been observed in modular proteins \cite{MyD12,fis00}
($N\sim 10$), whereas the rips forces are basically independent of $J$
for nucleic acids experiments (larger $N$)
\cite{rit06jpcm,hug10,HBFSByR10}.  Also, it is worth noting that it
has recently been shown that the length-controlled and the
flow-controlled scenarios in polymer stretching are thermodynamically
equivalent \cite{LHyS14}.

  Let us investigate in more detail the dependence of the force rips
  size with the number of units $N$ in the equilibrium case; see left
  panels of Fig.~\ref{force_rips}.  For large $N$, the free energy
  $\calA$ over each branch is extensive ($\propto N$), whereas the
  difference of free energies over consecutive branches for a given
  value of the length is independent of $N$. Therefore, the
  \textit{relative} free energy change between consecutive branches
  scales as $N^{-1}$. Making use of Eq.~\eqref{fe13} and neglecting
  terms of order $N^{-3}$, we obtain \cite{footnote3}
\begin{equation}\label{fe15}
 \frac{f_J^{\pm}-F_c\!}{\varphi_0}=\mp\frac{3\sqrt{3}}{N} \!\left(1\mp\frac{r_J}{N}\right)\!, \quad r_J\!=\left.\frac{2\calL_{J}}{\calL_N-\calL_0}\right|_{F_c}\! .
\end{equation}
Both $f_J^-$ and $f_J^+$ increase linearly with $J$, {and} so do $r_J$ and $\calL_J$, {[$\calL_J$ is given by eq. (\ref{fe8b})]}. This is necessary to fulfill the continuity condition for the free energy at the rips. On the other hand, for very large $N$, the term proportional to $r_J$ is {proportional} to $N^{-2}$ {and, therefore, it is small compared with} the first term on the rhs of eq.(\ref{fe15}), which is proportional to $N^{-1}$.  As a consequence, in this limit the force rips become independent of $J$ and symmetrical with respect to $F_c$,
\begin{equation}\label{fe15b}
  \frac{f_J^\pm-F_c}{\varphi_0}\sim \mp \frac{3\sqrt{3}}{N}.
\end{equation}
This is consistent with the behavior observed in nucleic acids
\cite{rit06jpcm,hug10,HBFSByR10}, in which the number of units is much
larger than that typical of modular proteins. Moreover, it shows that
the rip size in equilibrium follow a simple power law, it decays as
$N^{-1}$ for large $N$. We show the tendency to this power law
  in the bottom panel of Fig.~\ref{rips_size}, in which we plot the
  size of the rip $\Delta F=f_{J}^{-}-f_{J}^{+}$ for a specific value
  of $J$. We have chosen $J=(M+1)/2$, that is, the transition in which
  the number of unfolded units become larger than the number of folded
  ones, $J$ increases from $(M-1)/2$ to $(M+1)/2$ ($M$ odd).  Note
that the fact that $\lim_{N\to\infty}f_J^\pm=F_c$ implies that all the
units of the system unfold simultaneously at the critical force $F_c$
in the infinite size limit. This is the expected behavior, since in
the thermodynamic limit as $N\to\infty$ force fluctuations disappear
and the collectives with controlled force and controlled length should
be utterly equivalent.

\section{Chains with elastic interactions between identical modules}\label{real_chain}

In this Section, we investigate the effect of the harmonic potential
in Eq.~\eqref{1.1} (proportional to $(\eta_{j}-\eta_{j-1})^{2}$), on
the FECs. This term tends to minimize the number of ``domain walls''
separating regions with folded units from regions with unfolded units,
as the domain walls give a positive contribution to the free
  energy that is proportional to their number.  This elastic
interaction is expected to be more relevant in experiments in which
the unfolding/refolding of units is basically sequential, as
in the case of unzipping/rezipping of DNA/RNA hairpins. The
harmonic potential does not completely prevent the formation of
``bubbles'', regions of unfolded units inside a domain of folded ones,
but adds a free energy cost thereto. The same elastic
interaction is responsible for the so-called depinning
transition of wave fronts \cite{cb01,cba01,cb03,BT10}. The
  latter has been recently related to the experimentally observed
stepwise unfolding of modular proteins under force-clamp conditions
\cite{BCyP14}.

\subsection{Equilibrium states}\label{eq_real}

First, we consider the case in which there is no
disorder, all $k_j=k$ and $\delta_j=0$.  The equilibrium extensions
$\mathbf{\eta}^{\text{eq}}$ solve the minimization problem in
Eqs. (\ref{1.2c}) or (\ref{1.2fb}), that is, 
\begin{equation}\label{21.0}
 a'(\eta_j^\eq)-F+k(2 \eta_j^\eq-\eta_{j+1}^\eq-\eta_{j-1}^\eq)=0, \quad j=1,\ldots,N.
\end{equation}
These equations hold for all $j$, including the
  boundaries $1$, $N$, provided we introduce two fictitious extensions
  $\eta_{0}$, $\eta_{N+1}$, such that
\begin{eqnarray}
\eta_0=\eta_1, \quad \eta_{N+1}=\eta_N.
\label{3}
\end{eqnarray}
Alternatively, the extensions $\eta_j^\eq$ can be regarded as the
stationary solutions of the evolution equations (\ref{1}) with zero
noise. Again, in the length-controlled case, $F$ is a Lagrange
multiplier, calculated by imposing the constraint
$L=\calL(\bm{\eta})$. The equilibrium extensions may be found by
solving numerically \eqref{21.0}, but they can also be built
analytically by means of a perturbative expansion in powers of $k$, as
we now show.

\subsubsection{Pinned wave fronts for $k\ll 1$}\label{sec21}

Substituting the expansion
\begin{equation}\label{21.1}
  \eta_j^{\eq}=\sum_{n=0}^\infty \eta_{j,n}^{\eq} k^n, \quad j=1,\ldots,N,
\end{equation}
into Eq.~(\ref{21.0}), we obtain
\begin{subequations}
\begin{equation}\label{21.1a}
   a'(\eta_{j,0}^{\eq})= F,
\end{equation}
\begin{equation}\label{21.1b}
  \chi_j \eta_{j,1}^{\eq}=\eta_{j+1,0}^{\eq}+ \eta_{j-1,0}^{\eq}-2 \eta_{j,0}^{\eq},
\end{equation}
\begin{equation}\label{21.1c}
 \chi_j \eta_{j,2}^{\eq}= \eta_{j+1,1}^{\eq}+ \eta_{j-1,1}^{\eq}-2 \eta_{j,1}^{\eq}- \frac{1}{2}\zeta_j (\eta_{j,1}^{\eq})^2,
\end{equation}
\end{subequations}
where
\begin{equation}\label{21.2}
  \chi_j=a''(\eta_{j,0}^{\eq}), \quad \zeta_j=a'''(\eta_{j,0}^{\eq}).
\end{equation}

\begin{figure}[htbp]
\begin{center}
\includegraphics[width=3.25in]{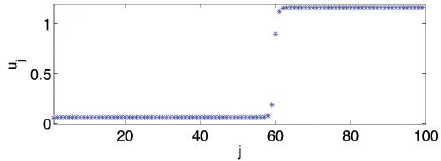}
\caption{
  Stable stationary wave
  front with increasing profile from $u^{(1)}$ to $u^{(3)}$
  (corresponding to $\1$ and $\3$, respectively) pinned at a
  particular point $j=J$ of an infinitely long chain, for $k=1.615$.
  The specular reflection of this pinned wave with respect to the
  center of the chain $j=N/2$ gives a pinned wave with decreasing
  profile from $\3$ to $\1$.  }
\label{fig1}
\end{center}
\end{figure}

For $k=0$ we recover the results of the previous section,
Eq. (\ref{21.1a}) is the same as Eq. (\ref{fe1b}).  The number of
``unfolded'' units $J$ having extensions $\3$ determines the
equilibrium values of Helmhotz free energy $\calA$, length $L$ and
Gibbs free energy $\calG$ of the considered configuration, as given by
eqs. (\ref{fe8a}), (\ref{fe8b}), and (\ref{fe9}), respectively. There
are $N!/[J!(N-J)!]$ configurations yielding the same values of $L$,
$\calA$, and $\calG$ for $k=0$, a degeneracy that is partially broken
at order $k$ by the elastic interaction. If three consecutive units
$(j-1,j,j+1)$ are in the same potential well (either folded or
unfolded) for $k=0$, then $\eta_{j,1}^{\eq}=0$ and the stationary
extension of the $j$-th unit does not vary. Therefore, only the
modules at the \textit{domain walls} separating domains where
$\eta_j=\1$ from others where $\eta_j=\3$ change their extension. At
the domain walls,
\begin{equation}\label{21.5}
  \eta_{j}^{\eq}=\left\{
  \begin{array}{l}
   \1+k\,\frac{\3-\1}{\chi^{(1)}}  +\mathcal{O}(k^2) \\
   \3-k\,\frac{\3-\1}{\chi^{(3)}}  +\mathcal{O}(k^2)
  \end{array}
  \right.
\end{equation}
The length of the folded (unfolded) unit is slightly increased (decreased), as observed
in Fig. \ref{fig1} for $k=1.615$. Therein, the second-order
  corrections in $k$ are already very small. Thus,
  in the remainder of this section, we neglect $\mathcal{O}(k^2)$
  terms, that is, we write all the expressions up to the linear
  corrections in $k$. The equilibrium length and free energy for $J$
unfolded units and $M$ domain walls are,
\begin{subequations}\label{21.6}
\begin{eqnarray}
  \calL_{J,M}^\eq &=& (N-J) \1 + J \3 \nonumber \\ && +  k \, M 
\frac{(\chi^{(3)}-\chi^{(1)})(\3-\1)}{\chi^{(1)}\chi^{(3)}}, \label{21.6a} \\
  \calG_{J,M}^\eq &=& (N-J) \guno+ J \gtres 
+k \, \frac{M}{2} (\3-\1)^2 . \nonumber \\
 \label{21.6b}
\end{eqnarray}
\end{subequations}
Thus, each domain wall contributes $k (\3-\1)
({\chi^{(1)}}^{-1}-{\chi^{(3)}}^{-1})$ to the length and $k
(\3-\1)^2/2$ to the free energy. An equivalent Ising model may
  be introduced to describe these equilibrium configurations, see
  Appendix \ref{ap_Ising}. The configurations with the
  fewest number of domain walls minimize the free energy $\calG$. For
the boundary conditions (\ref{3}), the minimal configurations have a
single domain wall for a given value of the number of unfolded
units $J$. The extension $\eta_j^{\text{eq}}$ increases with
$j$ from $\1$ to $\3$, slowly across the sites inside
either the folded and unfolded domains, and suddenly at the
domain wall, see Fig.~\ref{fig1}.

\subsubsection{Stability analysis}\label{sec:stability}

The \textit{pinned wave front} solutions in Fig.~\ref{fig1}
are stable in a certain range of forces, as proven in
the literature \cite{cb01,cba01,cb03,BT10}. Here, we investigate the
  stability for small $k$, by looking at the second variation of the
relevant thermodynamic potential.  We have
\begin{equation}\label{22.1}
\delta^2\calG= \delta^2\calA= \frac{1}{2}\sum_{j=1}^Na''(\eta_j^{\text{eq}})\, (\delta \eta_j)^2+\frac{k}{2}\sum_{j=1}^{N+1} (\delta \eta_j-\delta \eta_{j- 1})^2,
\end{equation}
where $\delta \eta_j=\eta_j-\eta_j^{\text{eq}}$.  Note that
the second variations of $A$ and $G$ are identical because
the term proportional to $F$ does not contribute to
  $\delta^{2}\calG$. It must be stressed that the
  non-diagonal terms of the symmetric matrix corresponding to this
  quadratic form are of order $k$, namely $\partial^2\calA/\partial
  \eta_j\partial \eta_{j\pm 1}=\partial^2\calG/\partial \eta_j\partial
  \eta_{j\pm 1}=-k$,  and they have not to be taken into
  account in our stability analysis.

\begin{figure}[htbp]
\begin{center}
\includegraphics[width=2.75in]{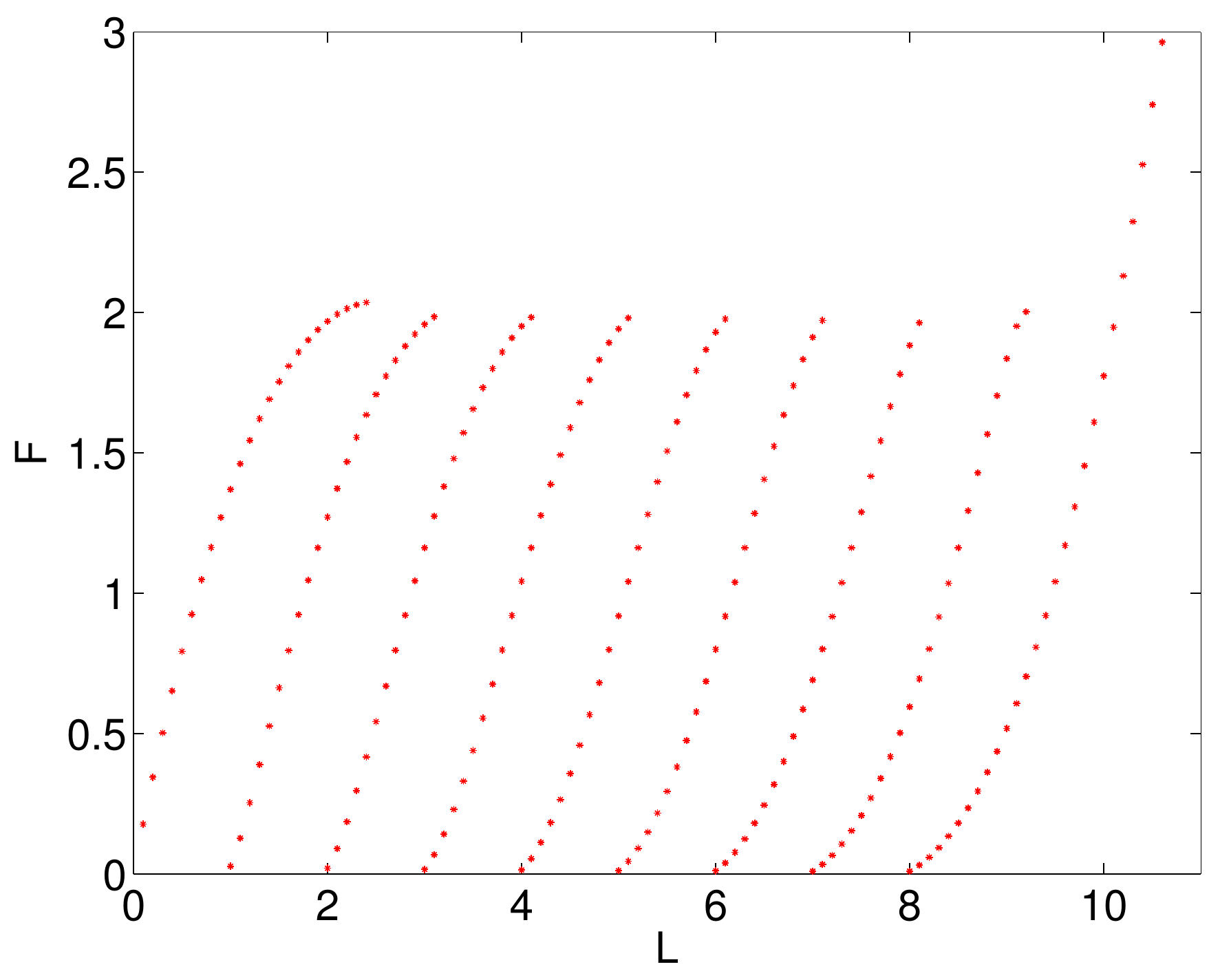}
\includegraphics[width=2.75in]{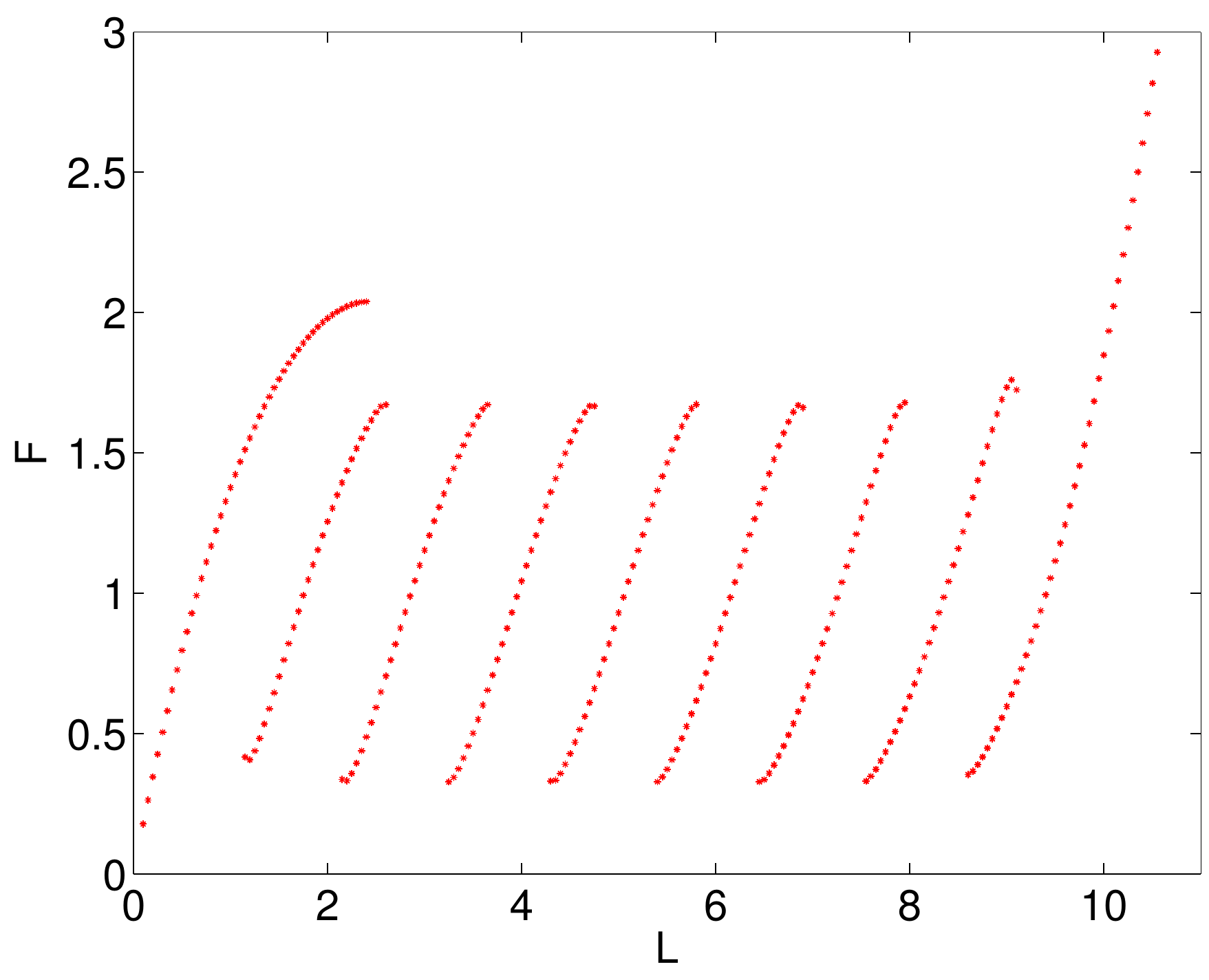}
\caption{FECs for a system with $N=8$ modules.
  (Top) Stable stationary branches for $k=0.055$, quite similar to
  those for $k=0$ (see Fig.~\ref{ramas_eq}). (Bottom) Stable
  stationary branches, each corresponding to a wave front pinned at a
  different site $j=J$, $J=1,\ldots,8$, for $k=0.55$. The completely
  folded and unfolded branches are basically unchanged, but the
  size of the intermediate branches is considerably reduced. Here $L$
  refers to the physical length \eqref{phys_length} that vanishes at $F=0$.}
\label{fig2}
\end{center}
\end{figure}

\begin{figure}[htbp]
\begin{center}
\includegraphics[width=3in]{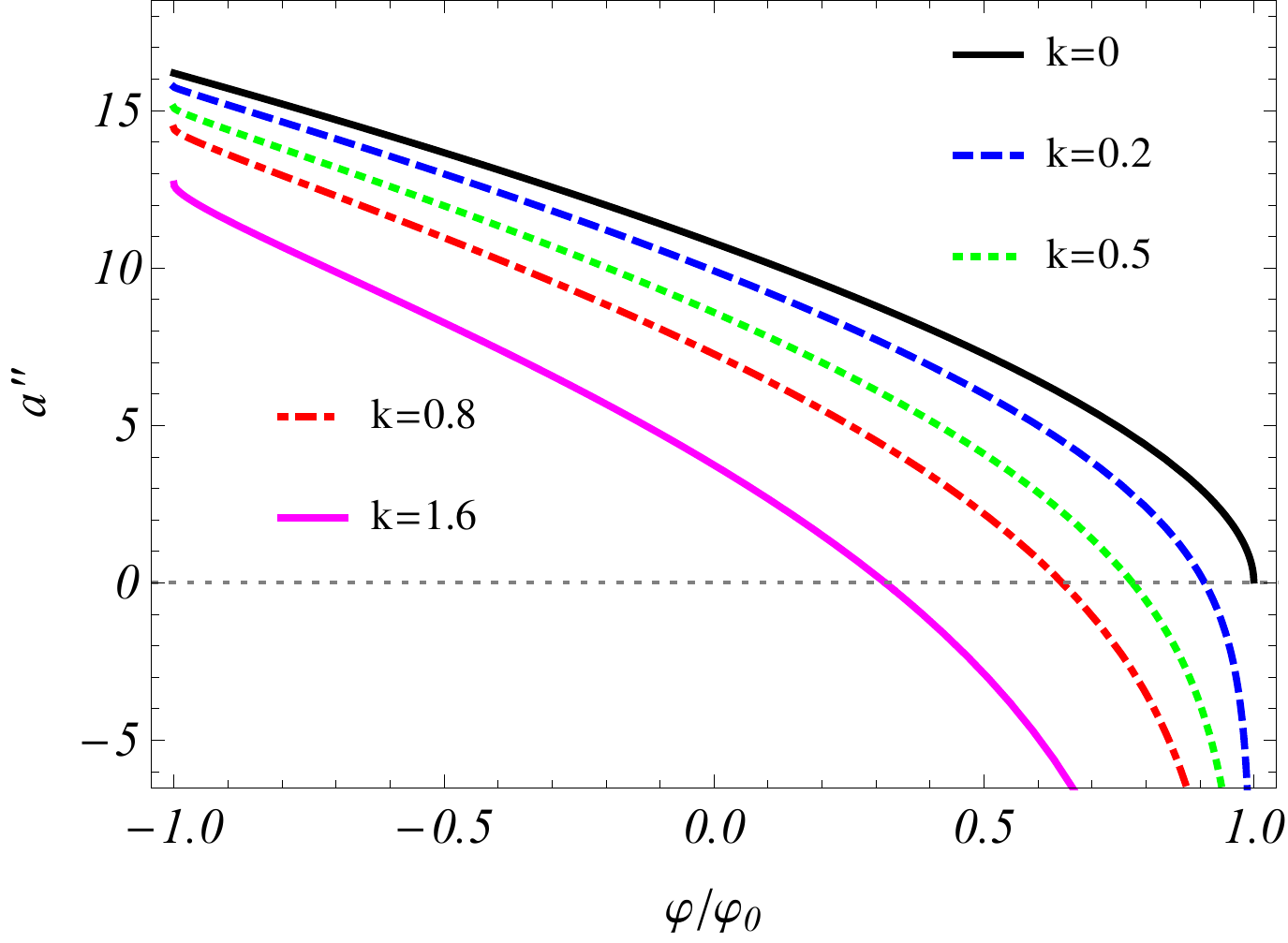}
\caption{Second derivative of the potential for the folded
  unit at the domain wall up to order $k$,
  Eq.~(\ref{22.4b}), as a function of the normalized  force
  $\varphi/\varphi_0$. It is clearly seen that the stable part of the
  branch, with $a''>0$, decreases with $k$. For $k=0.5$ the size of
  the branch is reduced by $20\%$, approximately, from its maximum
  size ($k=0$), consistently with the behavior observed in Fig.~\ref{fig2}.}
\label{stability}
\end{center}
\end{figure}

Let us consider a domain of folded (unfolded) units, whose
  lengths are $\eta^{(1)}$ ($\eta^{(3)}$) for the ideal chain with
  $k=0$.  Inside a domain of either folded or unfolded units, there is
  an additional positive contribution $2k$ to the diagonal terms
  $\partial^2\calA/\partial \eta_j^{2}$, so that stability is
  reinforced. Instability may arise at the domain walls, where 
\begin{equation}\label{22.4b}
\frac{\partial^{2}\calA}{\partial \eta_{j}^{2}}=\chi^{(i)}+k\left[2-\frac{|\zeta^{(i)}|(\3-\1)}{\chi^{(i)}}\right],
\quad  i=1,3.
\end{equation}
Consistently with the notation introduced in Eq.~\eqref{21.2},
$\zeta^{(i)}=a'''(\ei)=24\beta\ei$, $i=1,3$. Then
$\zeta^{(1)}<0<\zeta^{(3)}$ because $\1<0<\3$. The first and last branch of the FEC correspond to all-folded and to all-unfolded modules, respectively. Their configurations do not involve domain walls and therefore  $\partial^2\calA/\partial \eta_j^{2}=\chi^{(i)}$ for them, as in (\ref{22.4b}) with $k=0$. These branches are stable until $\chi^{(i)}=0$ at the extrema of $a'(\eta^{\text{eq}})$. In contrast to this, the other FEC branches have configurations with one domain wall and the linear
corrections in Eq.~\eqref{22.4b} cause $\partial^2\calA/\partial \eta_j^{2}$ to vanish for intermediate elongations between the extrema of $a'(\eta^{\text{eq}})$. As the limit of stability of the FEC branches is given by the condition
$\partial^{2}\calA/\partial \eta_{j}^{2}=0$, this reduces their size. This reduction in the branch size with $k$ is clearly observed in
  Fig.~\ref{fig2}. We further illustrate this result in
Fig.~\ref{stability}, where we plot the second derivatives of the
on-site potential at the domain wall, both for $k=0$ and with the
linear correction in $k$ (only for the folded unit
at the domain wall, the curves for the unfolded unit are just
the symmetrical ones with respect to $F_{c}=1$).

\subsection{Deterministic dynamics}\label{dynamics_T=0}

As the interacting chain is more complex than the ideal one, we start
by neglecting thermal noise. This corresponds to the so-called
deterministic (or macroscopic) approximation of the Langevin equation
\cite{vK92}. Alternatively, this can be presented as solving the
dynamical equations \eqref{langevin_eqs} at $T=0$.  In a later
Section, we will consider the changes introduced by a finite value of
the temperature.  Borrowing the usual terminology in classical
mechanics, we refer to slow processes at $T=0$ as \textit{adiabatic},
as they can no longer be regarded quasi-static because ergodicity is
broken.

\begin{figure}[htbp]
\begin{center}
\includegraphics[width=2.75in]{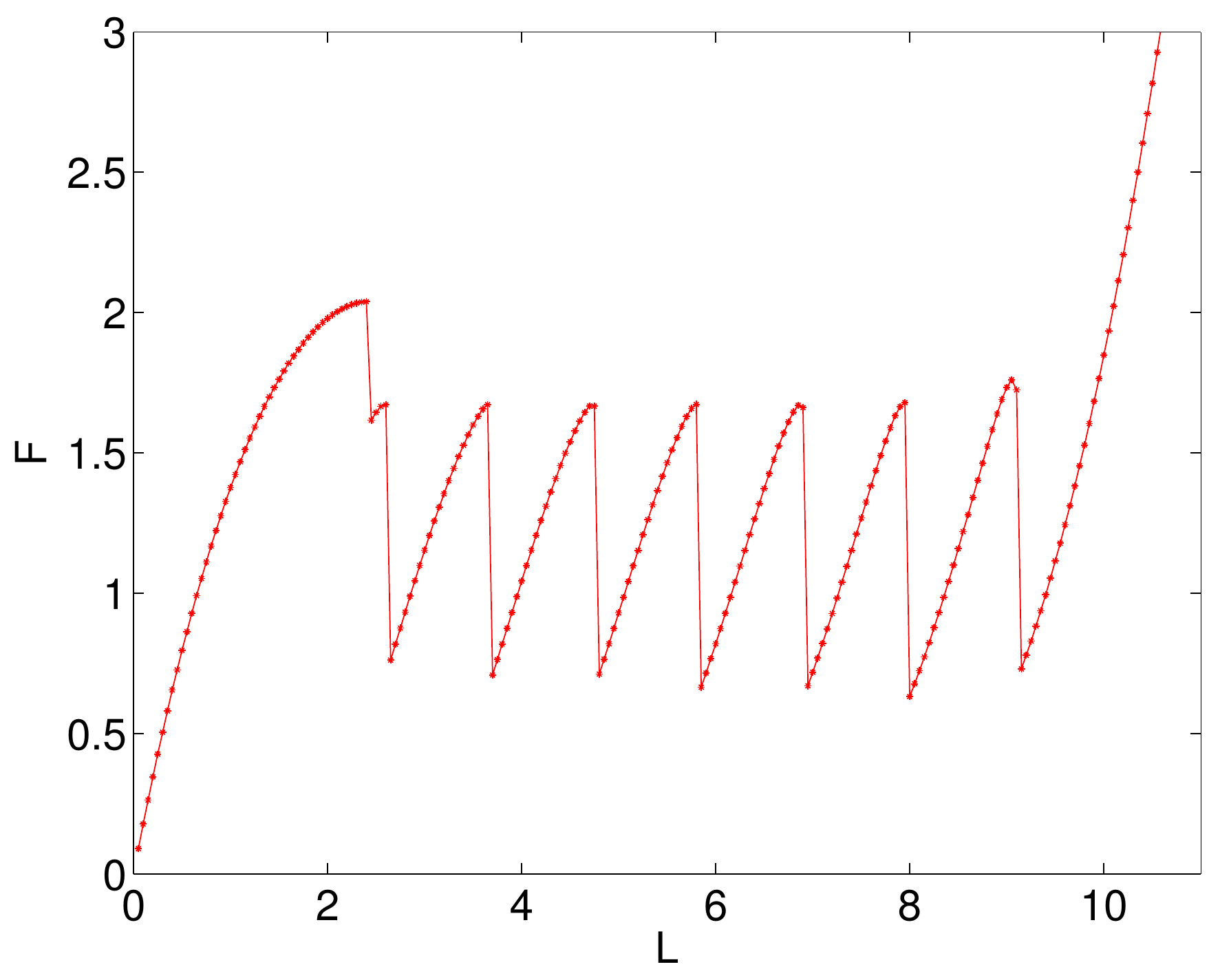}
\includegraphics[width=2.75in]{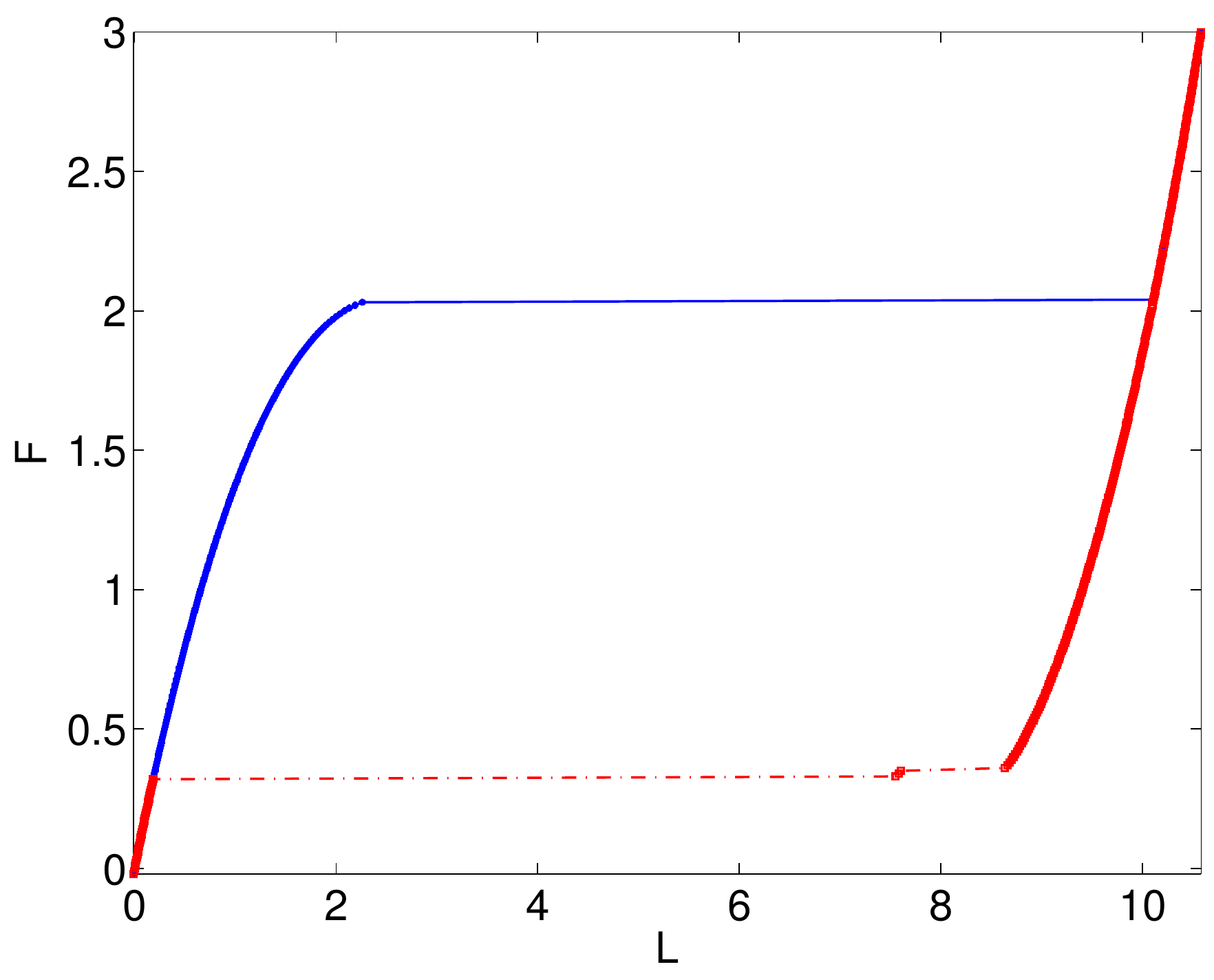}
\caption{(Top) FEC obtained by adiabatically increasing the length of the system. The local maxima of the branches are close to the corresponding upper ends of the equilibrium branches in Fig. \ref{fig2}(b). (Bottom) Hysteresis loop obtained by adiabatic force sweeping the force--extension diagram of a chain of identical units at zero temperature. In both cases, $k=0.55$. $L$ is the physical length that vanishes at $F=0$, defined in  Eq.~\eqref{phys_length}.}
\label{adiabaticT=0}
\end{center}
\end{figure}

In Fig. \ref{adiabaticT=0}, we plot two such processes. In the first
one (top panel) we increase the length adiabatically in a stepwise
manner, at each value of the length the system relaxes for a time
$\Delta t$, after which the length is increased in $\Delta L$. We have
chosen $\Delta L=0.2$ and $\Delta t=300$, for $k=0.5$. As compared to
the equilibrium branches in Fig. \ref{fig2}, we observe that the
$J$-th branch is swept as long it is locally stable, that is, until we
reach the maximum value of the force $F_{J,\text{max}}$ at which
$\delta^2\calA$ in (\ref{22.1}) is no longer positive definite. Then
the completely unfolded branch $J=0$ is swept to a higher force than
all the intermediate branches: Its size is not reduced with respect to
the $k=0$ case and $F_{0,\text{max}}>F_{J,\text{max}}$,
$J=1,\ldots,N-1$, as discussed in
Sec.~\ref{sec:stability}. This means that the portion of the
  $J=1$ branch that is swept is smaller than all of the other
  intermediate branches ($J\neq 0,N$) and there appears a ``bump'' in
  the FEC at the transition point between the $J=0$ and $J=1$ branch.
  This is clearly observed in the top panel of Fig. \ref{adiabaticT=0}
  around the length corresponding to the transition from the $J=0$ to
  the $J=1$ branch, $L\simeq 2.4$.  At this length, the corresponding
  force over the $J=1$ branch is much closer to the limit of stability
  of the intermediate branches than (for instance) the force over the
  $J=3$ branch at the transition length between the $J=2$ and $J=3$
  branches ($L\simeq 3.7$). In the force-controlled case (bottom
panel), we first increase the force adiabatically from $F=0$. The
system moves over the branch of folded units, $J=0$, until it reaches
the maximum thereof, $F_{0,\text{max}}$, at which the length jumps by
$\Delta L=N[\3(F_{0,\text{max}})-\1(F_{0,\text{max}})]$ to the
completely unfolded branch where $\eta_j=\3(F_{\text{max}})$ for all
$j$. If the force is now adiabatically decreased, the system moves
over the branch of unfolded units, $J=N$, until the force reaches its
minimum possible value and the system jumps back to the completely
folded branch. Thus, for both length-controlled and force-controlled
conditions, the largest possible hysteresis cycles appear, similar to
the ones obtained in storage systems, see Fig.~5 of Ref.~\cite{dre10}
or Fig.~7 of Ref.~\cite{dre11}.

\subsection{Influence of quenched disorder}
\label{quenched_disorder}

The biomolecules that are unfolded/refolded in the actual experiments,
nucleic acids and proteins, are actually heteropolymers, as the units
comprising a chain are not perfectly identical. This has led to
investigate the effect of their intrinsic quenched disorder (or,
equivalently, their intrinsic inhomogeneity) on their behavior in
different physical situations \cite{BDyM98,GyT06,AyA07}. In the
present context, their on-site double-well potentials $a(\eta_j)$,
their friction coefficients $\gamma_j$ and the spring constant between
modules $k_j$ may depend on $j$. These considerations are much more
important for DNA or RNA hairpins than for modular proteins. In the
latter, the units in our mesoscopic picture are the modules, which
have been artificially engineered to be as similar to each other as
possible. In the equivalent experiment to find the current-voltage
curves of superconductor superlattices, quenched disorder arises from
fluctuations of the doping density at different wells
\cite{GHMyP91,RTGyP02}. Including the natural variation in the free
energy parameters amounts to adding quenched noises to them.  To be
concrete, we consider a potential whose strength depends on a random
number $\delta_{j}$,
\begin{equation}
    a'(\eta_j;\delta_j)=(1+\delta_j)a'(\eta_j,\delta_j=0). \label{28}
\end{equation}
which are
i.i.d. random variables uniformly distributed on an interval
$[-\beta,\beta]$ ($\beta<1$).
\begin{figure}[htbp]
    \begin{center}
      \includegraphics[width=2.75in]{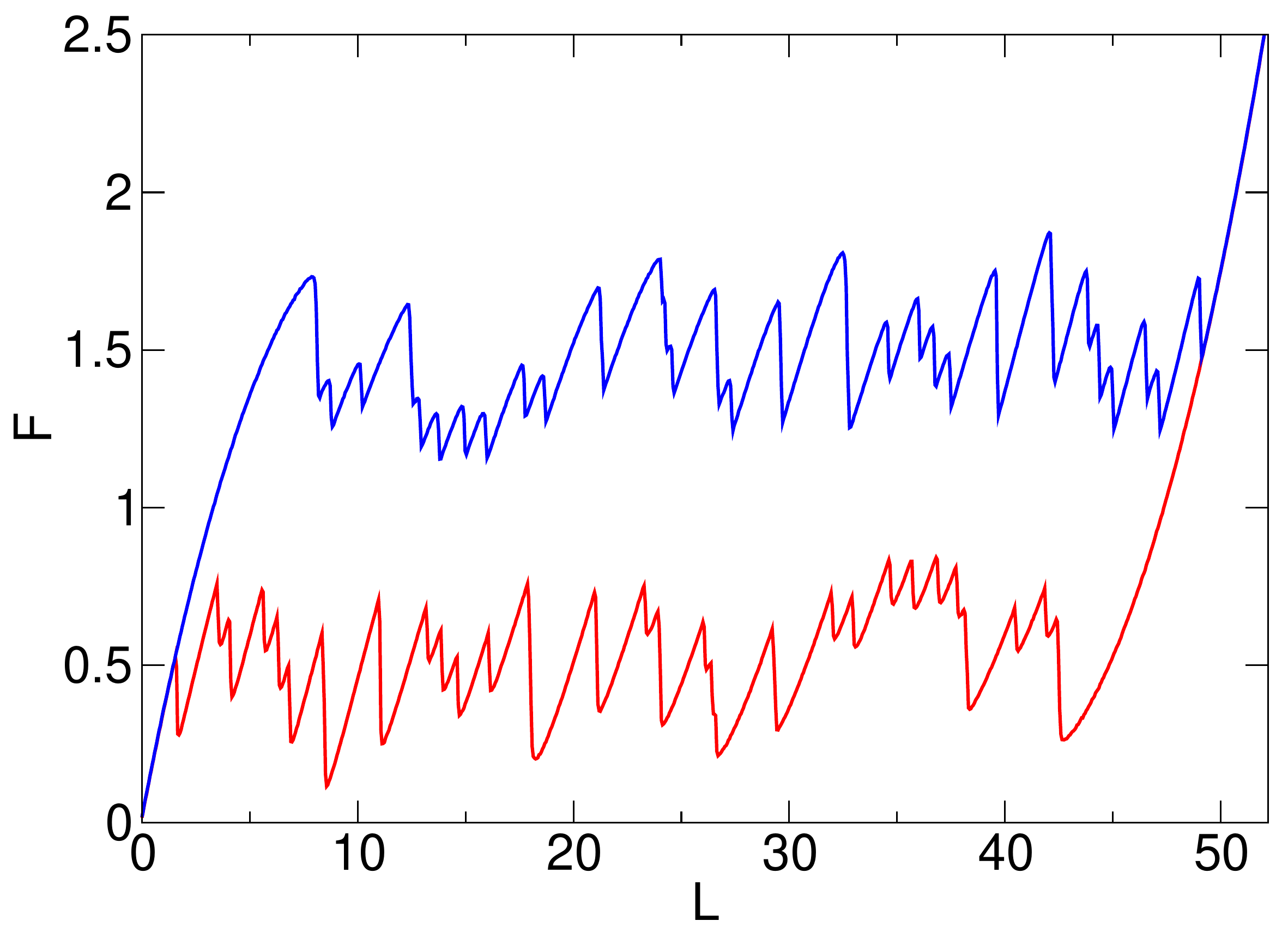}
      \includegraphics[width=2.75in]{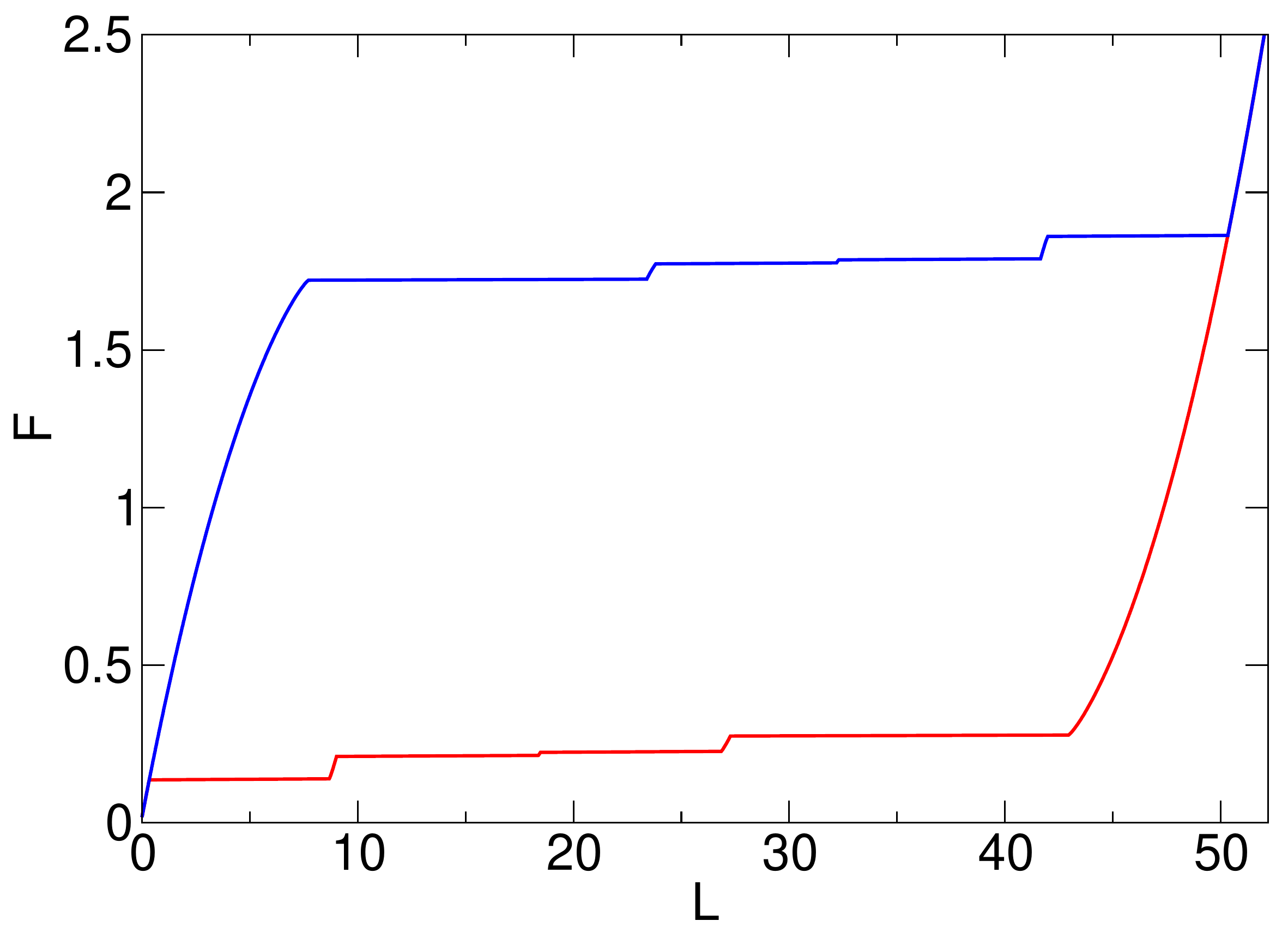}
      \caption{FEC for a DNA hairpin as in Figure
        \ref{fig1} but with $N=40$ and disorder as in Eq.~(\ref{28})
        (strength of the potential with $\beta=0.5$ and $k=1$).
        (Top) Hysteresis under length controlled
        conditions. The upper (lower) curve corresponds to
          adiabatically increasing (decreasing) length, with a rate
          $|dL/dt|=1.2\times 10^{-3}$ or smaller.
          (Bottom) Hysteresis under force-controlled
        conditions. Similarly, the upper (lower) curve
          corresponds to adiabatically increasing (decreasing) force,
          with a rate $|dF/dt|=3\times 10^{-3}$ or smaller. In the
        plots, $L$ is the physical length introduced in
        Eq.~\eqref{phys_length}. }
      \label{fig3}
    \end{center}
  \end{figure}

  Quenched disorder modifies both the stability of the FEC and the
  dynamics of the chain. When we depict the solutions corresponding to
  a wavefront pinned at particular locations as in Fig. \ref{fig2},
  the presence of disorder moves the solution branches up and down and
  affects the dynamical behavior of the system.  We show a hysteresis
  cycle under length-controlled conditions in the top panel
  of Figure \ref{fig3}. We have used a large disorder ($\beta=0.5$)
  which produces large variations in the length and height of the
  branches. Under force-controlled conditions, up and down
  sweeping the FEC, we obtain the much wider hysteresis cycles of the
  bottom panel of Fig. \ref{fig3}. Since the disorder changes the
  length and size of the force--extension branches, additional steps
  are seen in the hysteresis cycles, as compared to the case of identical units.

  \subsection{Influence of thermal noise}
  \label{thermal_noise}

In the last Section, we considered the effect of quenched disorder,
but we still had zero temperature.  Thermal noise allows random jumps
between stable branches, provided the system has sufficient waiting
time to escape the corresponding basins of attraction. As the control
parameter (force or length) changes more slowly, the  behavior of the system approaches the corresponding equilibrium statistical mechanics curve.

\begin{figure}[htbp]
    \begin{center}
      \includegraphics[width=2.75in]{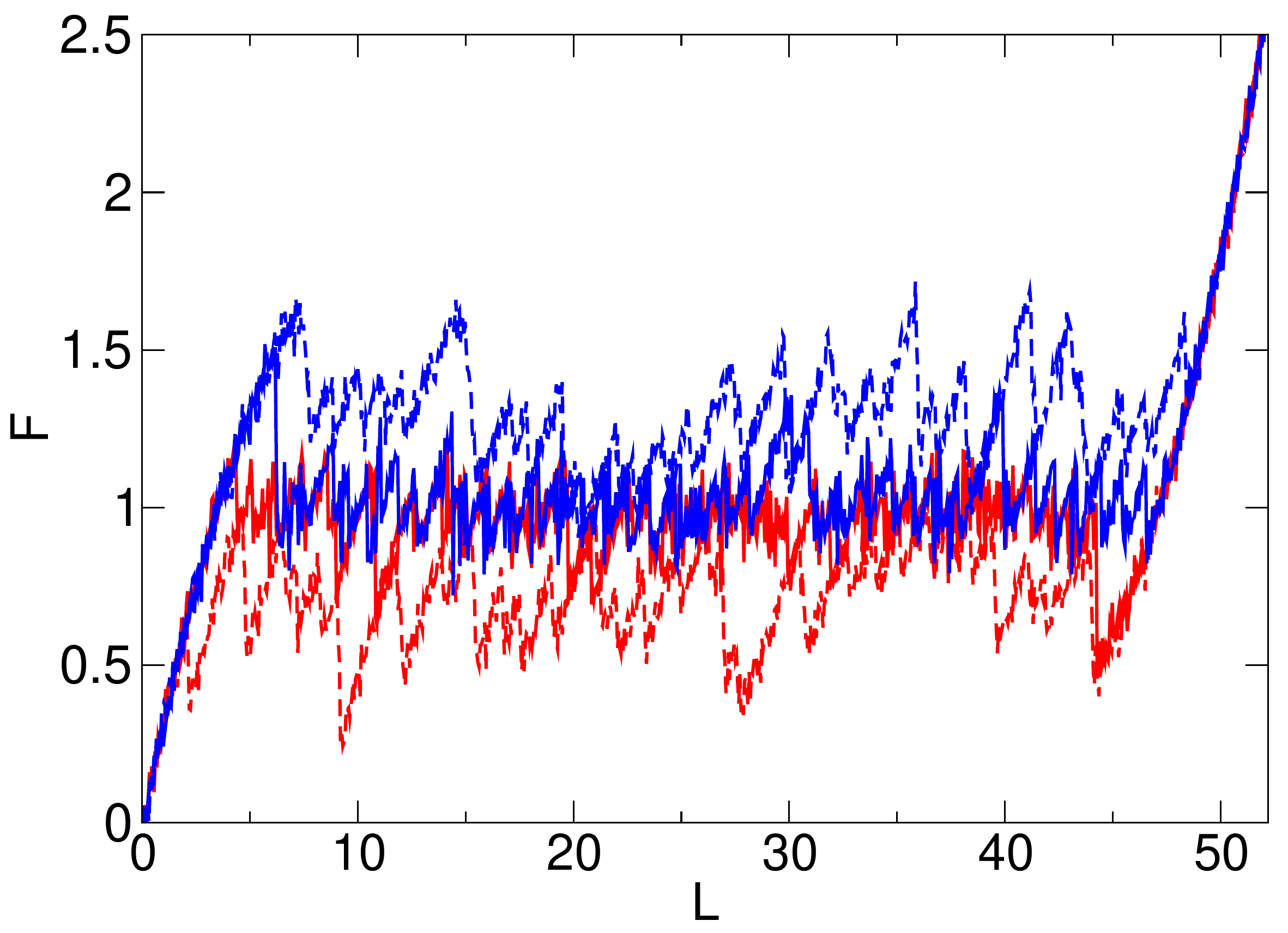}
      \includegraphics[width=2.75in]{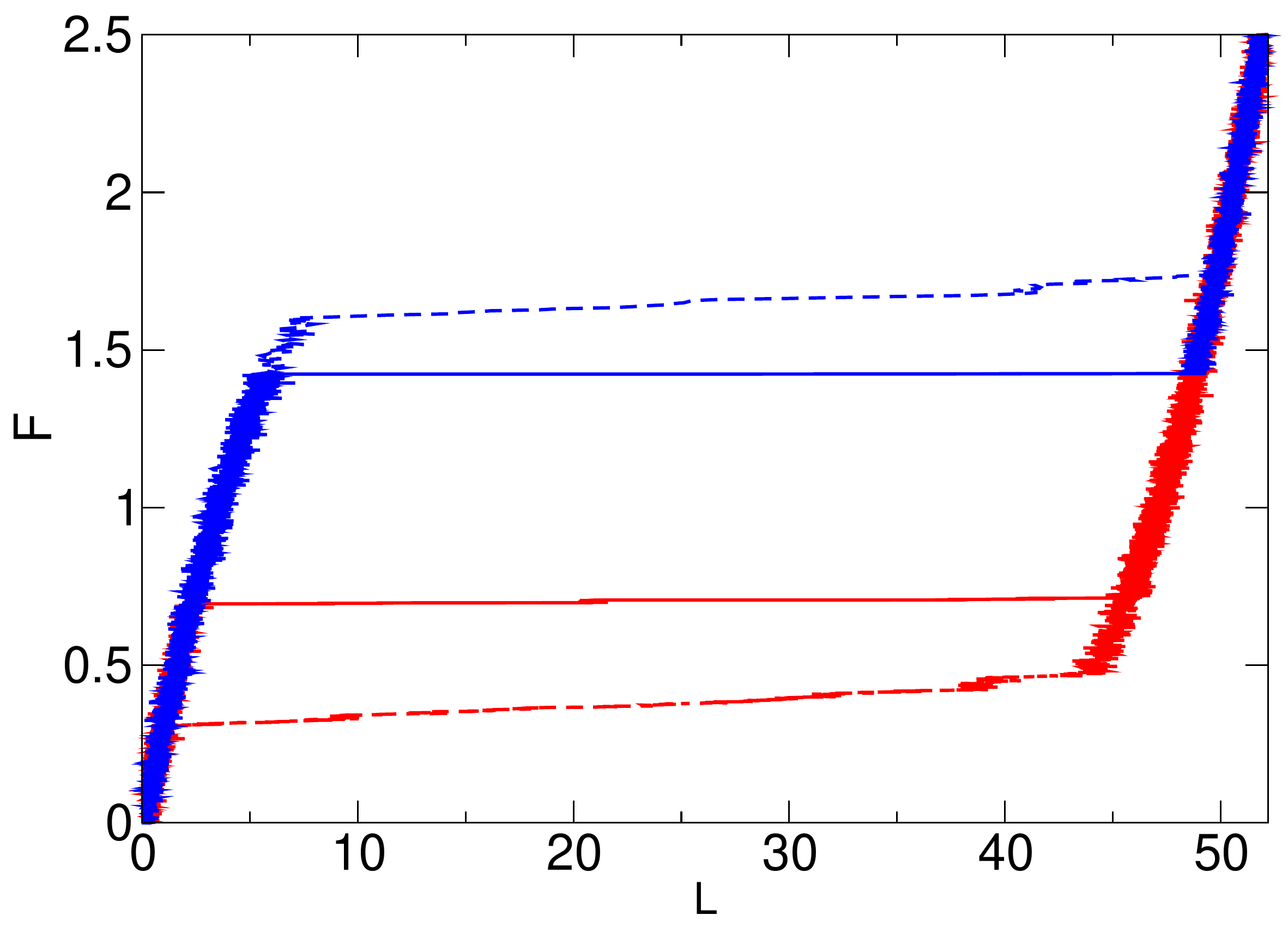}
      \caption{Same as in Fig.~\ref{fig3} but with additional white
noise of temperature $T=0.02$.  Thermal noise may suppress and
blur solution branches. In both panels, the dashed curves
correspond to the same rate as in Fig.~\ref{fig3}, while in the solid
ones the rate has been reduced by a factor $10^{-3}$. Again, $L$ is the physical length introduced
in \eqref{phys_length}.}
      \label{fig4}
    \end{center}
  \end{figure}

  Let us first consider length-controlled simulations. For an ideal
  biomolecule with identical modules, at $T=0$ adiabatic sweeping the
  FEC produces hysteresis cycles similar to the ones shown in
  Fig.~\ref{force_rips} for a very low temperature.  For finite
  temperature, (i) the size of the hysteresis cycles depends on the
  sweeping rate and becomes smaller as the rate decreases; (ii) there
  appear random jumps between stable branches that correspond to the
  same extension. Both effects have been observed in experiments with
  DNA hairpins, for which noise is much more important than in the
  case of modular proteins \cite{man05,lip01,lip02,hug10}. Also, some
  branches are not swept and the distinction between different
  branches is blurred, as shown in the top panel of
  Fig. \ref{fig4}. For a similar situation in semiconductor
  superlatices, see Fig.~2 in Ref.~\cite{RTGyP02}, which shows a
  current-voltage curve for a sample comprising $40$ periods of $9$nm
  wide GaAs wells and $4$nm wide AlAs barriers. It is also interesting
  to note that there is always some ``intrinsic'' hysteresis in the
  last (first) rip of the FEC, even for the lowest rate for which a
  perfect reversible behavior was obtained in the ideal case. This
  behavior has been observed experimentally in the unzipping/rezipping
  of DNA, see Fig.~1C and Fig.~S4 of Ref.~\cite{HBFSByR10}, and also
  in superlattices, see Fig.~1 of Ref.~\cite{GHMyP91} and Fig.~1 of
  Ref.~\cite{BHGyT12}. As explained in Section \ref{sec:stability},
  the FEC branch size is reduced in the non-ideal case ($k\neq 0$)
  except for the first and last branches whose configurations do not
  possess a domain wall. Then the non-zero interaction between
  neighboring modules makes the metastable regions in the first
  (completely folded) and the last (completely unfolded) branches
  wider than the rest.

In the force-controlled simulations, the effect of a finite
temperature is shown in the bottom panel of Fig.~\ref{fig4}. We
observe a behavior similar to that in Fig.~\ref{disc_L} for the ideal
chain, and also to the one found in other models \cite{Ka12,pra12}.
The physical picture is completely consistent with the experimental
findings in nucleic acids \cite{hug10}.

\section{Final remarks}
\label{conclusions}

We have proposed a biomolecule model that includes an on-site quartic
double-well potential and an elastic harmonic interaction among its
modules in the free energy thereof. Despite its simplicity, it
captures the main features of FECs in real biomolecules while allowing
us to identify the main physical mechanisms and to keep a
mathematically rigorous approach. This can be done in equilibrium but
also for the dynamics, for which we have written the
relevant Langevin (or Fokker-Planck equations). It should be stressed
that the Fokker-Planck equation for the length-controlled case is not
trivial, since the force $F$ appearing in the Langevin equation is an
unknown that must be calculated by imposing the length constraint. The
relevant thermodynamic potential, Gibbs-like (Helmholtz-like) for the
force-controlled (length-controlled) case, has been shown to be the
stationary solution of the Fokker-Planck equation.

Equilibrium FECs show multistability in a certain range of forces:
There are multiple FEC branches corresponding to different number of
folded/unfolded units. Under force-controlled conditions, there is an
equilibrium phase transition between the all-modules-folded to the
all-modules-unfolded, the lengths across the jump being determined by
continuity of force and Gibbs free energy. Under length-controlled
conditions, there appears a sawtooth FEC consisting of a number of
branches with force jumps between them in which the number of unfolded
modules differs by one.  The forces across the jump are determined by
continuity of length and Helmholtz free energy. In experiments, the
unfolding/refolding transitions take place neither at a perfectly
constant force nor at a perfectly constant length as seen in
Figs.~\ref{disc_L} and \ref{force_rips}, because of the finite
resolution of the the devices controlling the force or the
length. Thus, the controlled quantity is not exactly equal to the
desired value and also changes at the transition.

Dynamical FECs are obtained when the control parameter (either the
force $F$ or the length $L$) is changed at a finite rate: Some
hysteresis is present and the unfolding (refolding) forces increase
(decrease) with the rate, as observed in experiments
\cite{MNLRyE99,fis00,man05,lip01,lip02,fis00,hug10,CRJSTyB05}.  A
{crucial} role is played by the time {that} the system {needs} to
surpass the energy barrier regulating the transitions from the folded
to the unfolded state and viceversa. The key point is how this
Arrhenius time scale compares to that defined by the variation of the
force or length: It is only when the characteristic time defined by
the variation of the force or length is longer than the Arrhenius time
that the equilibrium FECs are recovered, because the force/length
program can {then} be considered quasi-static and there is no
hysteresis. We have shown in several cases throughout the paper that
this feature implies that a decrease in the temperature (while keeping
the rate of variation of the force/length) leads to a much wider
hysteresis cycle. In this cycle, the system typically sweep the whole
metastability region ($T=0$ or deterministic case).

Our results show that, in these elasticity experiments, biomolecules
display what may be called a ``metastable equilibrium behavior''. They
follow stationary FEC branches that can be obtained out of the
equilibrium solution of the Fokker-Planck equation, and dynamic
out-of-equilibrium excursions do not depart too much from them.  The
hysteresis cycles, completely similar to those observed in real
experiments, stem from equilibrium multistability: At the highest
loading rates, the system is not able to reach the absolute minimum
but sweeps a certain part of the metastable region (the narrower the
smaller the rate is) of the equilibrium free energy landscape. There
are techniques to obtain single molecule free energy differences from
time-dependent driving about hysteresis cycles
\cite{lip01,lip02,man05,CRJSTyB05}. In addition, the complete single
molecule free energy landscape can be obtained using model-dependent
algorithms \cite{HyS10}. Although there is some evidence of glass-like
behavior in force-clamp experiments with proteins \cite{BHWyF06},
hysteresis in these unfolding/refolding experiments seems to be quite
different from the more complex out-of-equilibrium hysteresis of
glassy systems in cooling/heating cycles. When cooled down to low
temperatures, glassy materials depart from the equilibrium curve and
end up in a far from equilibrium state; when reheated, they return to
equilibrium approaching a normal curve, which typically overshoots the
equilibrium one \cite{ByP93,BPyR94,PByS97,RyS03,KGyC13}.

We have also discussed in detail the role of the interaction between
neighboring units of the chain. The main effect of this interaction is
the reduction of the width of the  metastability region. When
the elastic interaction is absent, all the configurations with the
same number $J$ of unfolded units have the same free energy. This
``entropic term'' is reduced when the elastic interaction is taken
into account, since the free energy also depends on the number of
domain walls separating regions of unfolded and folded units and the
configuration with only one domain wall (pinned wave front) is
favored. From a physical point of view, this decrease is responsible
for the reduction of force fluctuations, which are at the root of the
width of the metastability region. Thus, we expect that the same
behavior will be present for more realistic interaction potential
between modules. In real biomolecules for which their on-site
potentials and number of modules are similar, a smaller size of the
rips may be linked to a stronger interaction between the neighboring
units. 

The relevance of the interaction between units is also clearly shown
by the fact that the metastability regions in the first (completely
folded) and last (completely unfolded) branches are wider than those
of the intermediate ones. This leads to the existence of some
``intrinsic'' hysteresis in the first/last force rips of the FEC under
length-controlled conditions, even for very low pulling rates, close
to the quasi-static limit. Interestingly, this effect has been
reported in experiments with DNA molecules, see for instance the FECs
in Fig.~1C (rezipping) and Fig.~S4 (unzipping) of
Ref.~\cite{HBFSByR10}.  In the unzipping (rezipping) experiment, the
physical reason is the ``extra'' free energy cost $k(\3-\1)^{2}/2$ for
creating (removing) the domain wall separating the folded and unfolded
regions of the molecule.  Thus, the presence or absence of intrinsic
hysteresis may be used to discriminate the importance of the coupling
between units in biomolecules or in other physical systems. For
instance, compare Fig.~1 of \cite{BHGyT12} (or of \cite{GHMyP91}) to
Fig.~2 of \cite{RTGyP02} for the current-voltage curve obtained in the
analogous experimental situation in semiconductor superlattices.

Many of the main characteristic behaviors observed here:
multistability (multiple branches for a certain region of parameters
like those in Fig.~\ref{ramas_eq}), the associated sawtooth FECs for
length-controlled experiments, hysteresis effects when the control
parameters are changed at a finite rate, etc. also occur in
quite different physical situations, such as many particle
storage systems \cite{dre10,dre11,dre11cmt} and weakly
coupled semiconductor superlattices
\cite{GHMyP91,RTGyP02,BGr05,BT10,BHGyT12}.  This analogy stems from
the following common feature: all these systems comprise a
number of similar bistable units whose individual states may be
determined by a long-range interaction introduced by a global
constraint (total charge \cite{dre10,dre11,dre11cmt}, fixed voltage
bias \cite{GHMyP91,RTGyP02,BGr05,BT10,BHGyT12}).  Of course,
fine-detail differences appear in the observed behavior in each
physical situation, depending on the relevance of non-ideal effects,
such as interactions among modules, quenched disorder, or
the thermal noise considered here. For instance, the maximum
size hysteresis cycles, basically identical to the deterministic case,
have been observed in Refs.~\cite{dre10,dre11} for storage systems. This seems to indicate a lesser relevance of fluctuations in the latter.

Voltage biased semiconductor superlattices are
definitely out-of-equilibrium systems: electrons are
continuously injected and extracted from contacts, and their behaviors
include time-periodic and chaotic oscillations besides hysteretic
behavior \cite{BGr05}. Nonlinear charge transport in superlattices
cannot be described with the free-energy scaffolding available for
biomolecules. Instead, discrete drift-diffusion models based on
sequential tunneling between neighboring quantum wells are used
\cite{BGr05,BT10}. Nevertheless, the present paper shows that the
methodology developed for these discrete systems can be adapted to
describe FECs of biomolecules. As experiments with semiconductor
superlattices are much more controllable than those with biomolecules,
it would be interesting to see what the interpretation of measurements
given in Refs.~\cite{lip01,lip02,man05,CRJSTyB05,HyS10} produces in
the superlattice case.

According to the above discussion, our main conclusions are quite
general. They are applicable not only to biomolecules but to any
physical system composed of repeated similar bistable units. Of
course, we need renaming appropriately variables for each relevant
physical situation. For instance, force-extension curves must be
replaced by chemical potential-charge ones in storage systems
\cite{dre10,dre11} or by current-voltage curves in semiconductor
superlatices \cite{BGr05,BT10}. Depending on the system, some of the necessary experiments are not yet available.  For instance, there are no precise current-controlled experiments on semiconductor superlattices. Thus our investigations open new interesting perspectives for experimental research in these fields.

\acknowledgments
This work has been supported by the Spanish Ministerio de Econom\'\i a y Competitividad grants FIS2011-28838-C02-01 (LLB),  FIS2011-28838-C02-02 (AC), FIS2011-24460 (AP). 

\appendix

\section{Asymmetric realistic potential}\label{app:Berko}

Here we consider the effects of using a more realistic free energy for
the modules. This energy was first considered by Berkovich,
Garcia-Manyes, Klafter, Urbakh and Fernandez (BGMKUF) to model the
unfolding of single-unit proteins, such as I27 or ubiquitin, observed
in AFM experiments \cite{ber10}. Very recently, we have
  employed it to investigate stepwise unfolding of polyproteins
under force-clamp conditions \cite{BCyP14}. At zero force,
the BGMKUF potential for one unit is
\begin{eqnarray}
a(\eta) & = & U_0\!\left[\!\left(1-e^{-2b(\eta-R_c)/R_c}\right)^2-1\right]\! \nonumber\\&&
+\frac{k_BTL_c}{4P}\!\left(\frac{1}{1-\frac{\eta}{L_c}}-1-\frac{\eta}{L_c}
+\frac{2\eta^2}{L_c^2}\right)\!.  \label{eq:berkovich}
\end{eqnarray}
This free energy for each unit is the sum of an enthalpic contribution
given by a Morse potential and an entropic contribution given by a WLC
potential, \cite{ber10,ber12}. Under application of force, the energy
$A(x)-Fx$ exhibits two minima separated by a force-generated barrier
\cite{ber10}. Manifestations of the sensitive dependance of unfolding
and refolding on the barrier created by the applied force have been
experimentally measured in \cite{elm12,bai12}. Here we use the
parameter values of Ref.~\cite{ber10} (slightly different from those
in Ref.~\cite{BCyP14}), $P=0.4$nm (persistence length), $L_c=30$nm
(contour length), $T=300$K, $U_0=100$pN nm($\sim\!\!  24 k_B T$),
$R_c=4$nm, $b=2$. Force and extensions are measured in units of
$[F]=100$ pN and $L_c=30$ nm, respectively. We define dimensionless
variables, $\mu=U_0/(L_c[F])$, $\beta=2bL_c/R_c$, $\rho=R_c/L_c$,
$A=k_B TL_c/(4PU_0) $, thereby obtaining the following dimensionless
potential \begin{eqnarray}
            a(\eta)&=& \mu\!\left\{\left[1-e^{-\beta(\eta-\rho)}\right]^2-1 \right. \nonumber \\
                && \left.\quad +A\!\left(\frac{1}{1-\eta}-1-\eta+2\eta^2\right)
                   \right\}, \label{eq:berk-dimless}
\end{eqnarray}
As repeatedly done throughout the paper, we keep the same notation for
dimensionless and dimensional potentials. The dimensionless parameter
values in (\ref{eq:berk-dimless}) are $\mu=0.0333$, $\beta=30$,
$\rho=0.133$, and $A=0.776$.  On the other hand, the friction
coefficient $\gamma$, given by the Einstein relation $D=k_B T/\gamma$,
sets the time unit $[t]= \gamma L_c/[F]$. The diffusion coefficient
for tethered proteins in solution $D$ has a typical value $D =1500$
nm$^2$/s \cite{ber10}, so that $\gamma=0.00278$pN nm$^{-1}$s and
$[t]=0.833$ms.

\begin{figure}
\begin{center}
    \includegraphics[width=3.25in]{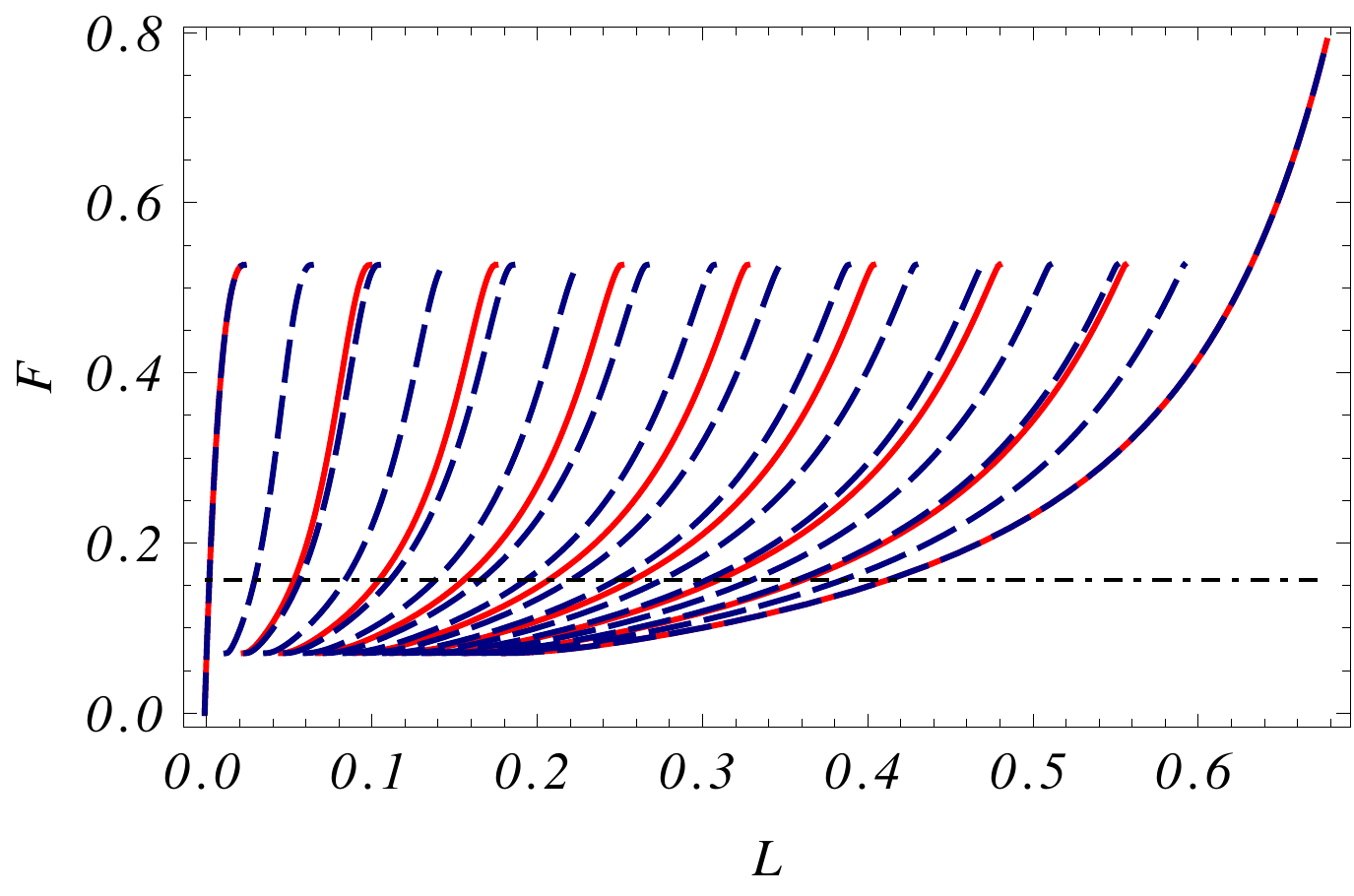}
    \caption{FECs for the BGMKUF potential, with $N=8$ (solid red)
      and $N=15$ (dashed blue).  There are $N+1$ branches in the
      metastability region $F_{m}<F<F_{M}$, with the number of
      unfolded units $J$ increasing from left to right.  The first
      ($J=0$) and last ($J=N$) branches are independent of $N$, as in
      Fig.~\ref{ramas_eq}.  Note the asymmetry of the branches
      with respect to the critical force $F_{c}$ (dot-dashed line).  }
    \label{fig:berko1}
\end{center}
\end{figure}

For this choice of parameters, there is metastability for forces in
the range $F_{m}<F<F_{M}$, with $F_{m}=0.704$ ($7.04$pN) and
$F_{M}=0.527$ ($52.7$pN). We show the equilibrium branches for two
systems, with $N=8$ and $N=15$, respectively, in
Fig.~\ref{fig:berko1}. Analogously to what we observed for the simple
Landau-like free energy in Fig.~\ref{ramas_eq}, the branches become
denser as the number of units increase. On the other hand, there is no
up-down {nor left-right} symmetry: The branches are no longer
symmetric with respect to {either} the critical force
$F_{c}=15.6$pN, at which the folded and unfolded minima are equally
deep, {or the central branch with half of the units unfolded,
  $J=N/2$}.  Here, and throughout this Section, we have considered the
``physical'' length corresponding to the extension of the molecule
with respect to its equilibrium length for zero force.

\begin{figure}
\begin{center}
    \includegraphics[width=3.25in]{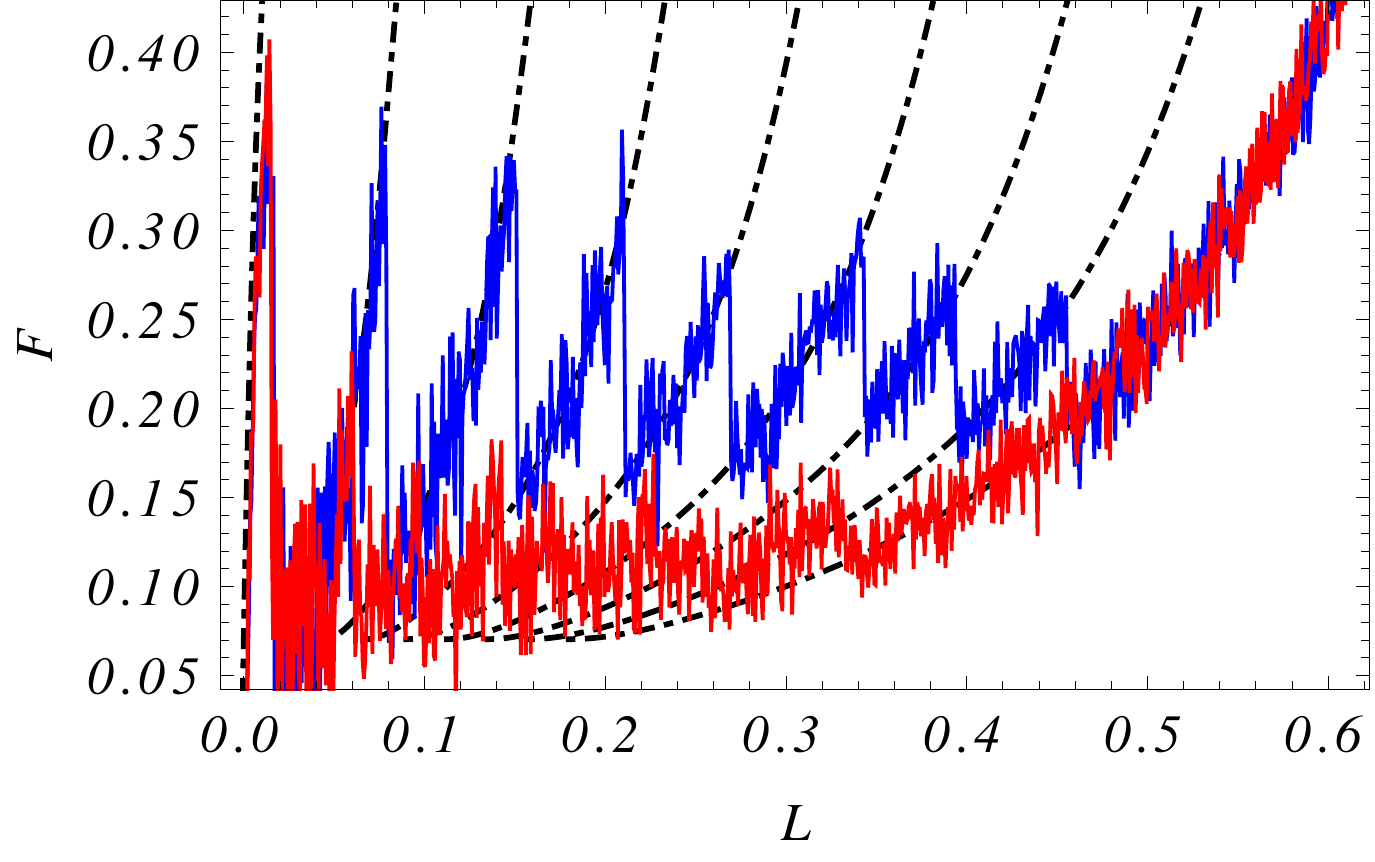}
    \caption{Unfolding/refolding cycle for a modular protein with $8$
      units with free energies given by the BGMKUF potential. The
      dot-dashed lines correspond to the equilibrium branches of the
      FEC, see Fig.~\ref{fig:berko1}. There are sharp force rips in
      the unfolding process, each corresponding to the unfolding of
      one of the units (jumps between neighboring branches). In the
      refolding process, there are no sharp peaks until the length has
      almost completely relaxed, $L\lesssim 0.2$.  }
    \label{fig:berko2}
\end{center}
\end{figure}

An unfolding/refolding cycle is shown in Fig.~\ref{fig:berko2}, in
which the length $L$ is increased at a constant rate. We show the FEC
corresponding to a typical AFM rate, namely $50$nm/s. In dimensionless
variables, this means that $\dot{L}=0.0014$, because the unit of
velocity is $L_{c}/[t]=3.6\times 10^{4}$nm/s. As for the Landau-like
potential considered in the text, the force has been averaged over a
certain time interval $\Delta t=1.2$ (corresponding to $1$ms) to mimic
the finite resolution of the measuring devices in real
experiments. Due to the asymmetry of the equilibrium branches, the
unfolding and refolding curves are quite different, as seen in
experiments. A clear sawtooth pattern is present in the unfolding
curve: The molecule clearly sweeps a certain part of each equilibrium
branch until it reaches a length at which it jumps to the neighboring
branch. {Similarly to experimental observations}, this jump is
associated to a decrease in the force (force rip)
\cite{HyD12,MyD12,fis00,car99}. On the other hand, in the refolding
process, the curve is much smoother and it is much more difficult to
identify the intermediate branch that the system is sweeping, at least
for the first stage of the relaxation curve (here, for
$L\gtrsim 0.2$). This is analogous to the usual experimental behavior in
the refolding process \cite{OMCyF99,RPSyG99,SSHyR02}.  However, there
appear clearer traces of force peaks in the refolding FEC when the
molecule has partially relaxed ($L\lesssim 0.2$). This behavior
resembles the  FECs obtained for the NI6C protein in
Ref.~\cite{lee10}, see Figs.~1C, 1D, and S5 therein.

Although the previous unfolding/refolding cycle is very similar to
those observed in experiments, it may be argued that our ideal
length-control device may have some impact on the observed
behavior. Therefore, we consider now a more realistic length-control
device, as the one depicted in Fig.~1 of Ref.~\cite{MyD12}, which
leads to the length-control potential term in
Eq.~\eqref{eq:length-control}, where $\chi_{\text{lc}}$ is the
(finite) spring constant of the cantilever. A typical value of the
spring constant for an AFM experiment is $6$pN/nm, which gives a
dimensionless value $\chi_{\text{lc}}=1.8$. Firstly, it is important
to stress that the equilibrium branches of the FEC are not changed by
the finite stiffness of the length-controlling device. The equilibrium
extensions $\eta_{i}$ are given by Eq.~\eqref{fe1b}, $a'(\eta_{i})=F$,
but now $F=-\chi_{\text{lc}}[L(\bm{\eta})-L]$ is the force exerted by
the finite stiffness control device. {Metastability appears} in the
same range of applied forces {as in the case of ideal length control,}
the only difference is that the end to end distance $L(\bm{\eta})$
does not equal $L$, {instead,}
$L(\bm{\eta})=L-F/\chi_{\text{lc}}<L$. In other words, the tip of
cantilever has an equilibrium deflection $\Delta x=F/\xi_{\text{lc}}$
for each considered force $F$.

\begin{figure}
\begin{center}
    \includegraphics[width=3.25in]{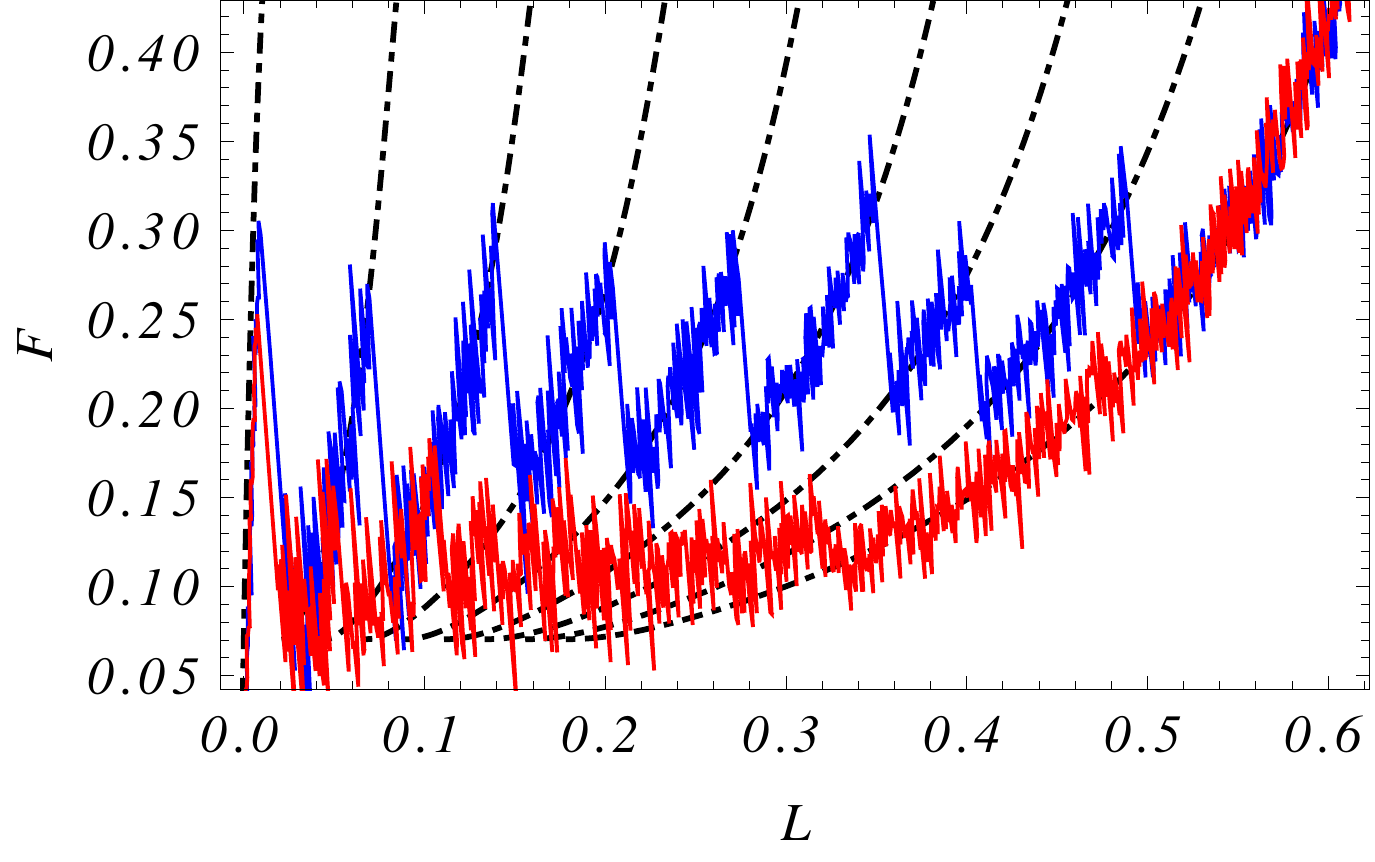}
    \caption{Unfolding/refolding cycle for a modular protein with $8$
      units with free energies given by the BGMKUF potential and a
      finite value of the cantilever stiffness. The force $F$ is
      plotted against the end-to-end distance of the molecule
      $L=L(\bm{\eta})$. The equilibrium branches of the FEC, see
      Fig.~\ref{fig:berko1}, are the dot-dashed lines. Both the
      unfolding and refolding curves are very similar to those in
      Fig.~\ref{fig:berko2}, except for the force rips in the
      unfolding process not being perfectly vertical as a consequence
      of the imperfect length control.  }
    \label{fig:berko3}
\end{center}
\end{figure}

 Repeating the
unfolding/refolding process in Fig.~\ref{fig:berko2}, with the only
difference of the finite value of the stiffness, we have obtained the
results shown in Fig.~\ref{fig:berko3}. The unfolding/refolding cycles
in both figures are very similar, although they would not match
perfectly when superimposed. To obtain complete agreement with the
perfect length control situation shown in Fig.~\ref{fig:berko2}, we
should have employed a larger value of the spring constant, around
$150$pN/nm or $\chi_{\text{lc}}=45$. In particular, the refolding
curve is again much smoother than the unfolding sawtooth pattern found
with either the quartic or the BGMKUF potential,
but with some minor upward traces for $L\lesssim 0.2$.

\section{Equivalent Ising model for the free energy minima \label{ap_Ising}}

We can write down the length and Gibbs free energy
(\ref{21.6}) in an Ising-like manner. Let us assign a
spin-down variable to the folded units, so
that $\sigma_j=-1$ if $\eta_{j,0}^{\eq}=\1$, and an
spin-up $\sigma_j=+1$ to the unfolded ones, with
$\eta_{j,0}^{\eq}=\3$. The number of unfolded units and
domain walls are
\begin{equation}\label{21.7}
  J=\sum_{j=1}^N \frac{1+\sigma_j}{2}, \qquad M=\sum_{j=1}^{N-1} \frac{1-\sigma_j \sigma_{j+1}}{2}.
\end{equation}
Except for an additive constant, the  free energy (\ref{21.6b}) becomes
\begin{equation}\label{21.8}
  \calG^{\eq}(\bm{\sigma})=-H \sum_{j=1}^N \sigma_j - \Xi \sum_{j=1}^{N-1} \sigma_j \sigma_{j+1}+\mathcal{O}(k^2),
\end{equation}
an Ising system with an external field $H$ and ferromagnetic nearest neighbor coupling $\Xi$ given by
\begin{equation}\label{21.9}
  H= \frac{\gtres-\guno}{2}, \quad \Xi=\frac{k[\3-\1]^2}{4}>0.
\end{equation}
Interestingly, a similar expression for the free energy was
  proposed in Ref.~\cite{makharov}. The sign of $H$ determines which
minimum of the Gibbs free energy $g(\eta)$ is deepest, $\1$
or $\3$; at the critical force $F_c=1$, that is, $H=0$, they are
equally deep. The ferromagnetic coupling $\Xi\propto k$ favors the
configurations with domains of parallel spins and thus a
minimal number of domain walls for a given number of unfolded
units $J$ \cite{note2}.  Then $M=0$, when all the
units are either folded or unfolded, or $M=1$, when there are
both folded and undolded units, produce the minimum free
energy (\ref{21.8}).

Given (\ref{21.7}), the length of the system at equilibrium is
\begin{eqnarray}\label{21.10}
  \calL^{\eq}(\bm{\sigma}) &=& \frac{N}{2} \left(\1 +\3\right)
  +(N-1)\Delta \nonumber \\
&&  +\frac{\3-\1}{2} \sum_{j=1}^N \sigma_j- \Delta \sum_{j=1}^{N-1} \sigma_j \sigma_{j+1} ,
\end{eqnarray}
where
\begin{equation}\label{21.11}
  \Delta= \frac{k[\chi^{(3)}-\chi^{(1)}]}{2\chi^{(1)}\chi^{(3)}}.
\end{equation}
The parameter $\Delta$ can be positive or negative. For the simple quartic potential we are considering, $\Delta=0$ at the critical force $F_c=1$, $\Delta>0$ for $F<F_c$, and $\Delta <0$ for $F>F_c$.

We have not considered here the quadratic
corrections, proportional to $k^2$, which only affect sites
at the domain walls and their nearest neighbors. In this
equivalent Ising description, they (i) change the first order
coupling constants $\Xi$ and $\Delta$, and (ii) introduce a
second-nearest-neighbor interaction. Similarly, by taking into
  account higher order corrections, up to order $k^n$, we get an
Ising model with longer-ranged interactions up to the
$n$th-nearest-neighbors.

\section{Lyapunov function for the deterministic dynamics in the
  length-controlled case \label{ap_Lyapunov}}

Unlike the Gibbs free energy $\calG$ in the force-controlled case, the
Helmholtz free energy $\calA$, as given by Eq.~\eqref{1.1},
is no longer a Lyapunov function of the zero-noise dynamics under
length-controlled conditions with a known length dependence
$L(t)$. However,
\begin{eqnarray}
\tilde{\calA}(\bm{\eta})&=&\calA(\bm{\eta})+\sum_{j=1}^N\!\Bigg[
\frac{k}{2}(\eta_{j+1}-\eta_{j})^2 \nonumber  \\ &&
-\frac{\eta_j}{N}\!\left(\sum_{k=1}^Na'(\eta_k)+\frac{dL}{dt}\!\right)\!-\frac{a(\eta_j)-\eta_ja'(\eta_j)}{N}\!\Bigg]\!. \nonumber\\
\label{mod_lyap}
\end{eqnarray}
is a Lyapunov function in this case. In fact, the governing nondimensional equations can be written as
  \begin{equation}
    \frac{d\eta_j}{dt}=-\frac{\partial}{\partial\eta_j}\tilde{\calA}(\bm{\eta}), 
    \label{zeroT_eq}
  \end{equation}
  after eliminating $F$ by means of Eq.~\eqref{1.3c}. Then 
$$\frac{d}{dt}\tilde{\calA}(\bm{\eta})=-\sum_{j=1}^N\!\left[\frac{\partial}{\partial\eta_j}\tilde{\calA}(\bm{\eta})\right]^2\!\leq 0.$$ 
  Also, 
\begin{eqnarray}
\tilde{\calA}(\bm{\eta})>N\min_{u}\!\left[a(u)-Fu-\frac{a(u)-ua'(u)}{N}\right]\!\nonumber
\end{eqnarray} 
for $$F_m<F=\frac{1}{N}\sum_{j=1}^N a'(\eta_j)+\frac{1}{N}\frac{dL}{dt}<F_M.$$


\begin{thebibliography}{99}
\bibitem{length} Thus, the total length should not be confused
    with the contour length of the biomolecule, which is a fixed
    quantity.
\bibitem{footnote1}
The length $L$ is extensive in the sense that it is proportional to the number of units $N$ in the limit as $N\gg 1$.
\bibitem{note} 
In biomolecules at zero force, the folded state is much more localized
than the unfolded state but this feature cannot be reproduced with the
simple Landau-like potential we are using. Due to its simplicity, the barrier separating the two minima is always
closer to the metastable state. In order to make the ``width'' of the
folded state smaller than that of the unfolded state, more realistic
potentials like the one in Refs.~\cite{ber10,ber12,BCyP14} must be used.
\bibitem{footnote2} This time interval is much longer than the time
  step used to integrate the Langevin equations, but much shorter than
  the total time for the stretching (or relaxing) process. As
  a result, the plotted force is close to $F_{\text{exp}}$,
  Eq.~\eqref{eq:Fexp}, because the time average of $\Delta F$ is very
  small.
\bibitem{footnote3} In Ref.~\cite{pra13}, it was $L_{J-1}+L_{J}$
  instead of $2L_{J}$ in the numerator of $r_{J}$. Equation
  \eqref{fe15} is more consistent, since $L_{J-1}-L_{J}$ is of the
  order of $N^{-1}$.
\bibitem{note2}
The boundary conditions \eqref{3} imply that
the terms corresponding to $j=0$ and $j=N$ do not contribute to the
interaction between nearest neighbor spins.
\end{thebibliography}
\end{document}